\documentclass[twocolumn]{aastex631}

\usepackage{multirow}
\usepackage{chngcntr}
\received{November 24, 2021}
\accepted{March 3, 2022}
\submitjournal{ApJL}

\shorttitle{SVS13-A}
\shortauthors{Bianchi et al.}

\begin{document}
\title{The two hot corinos of the SVS13-A protostellar binary system: counterposed siblings}


\correspondingauthor{Eleonora Bianchi}
\email{eleonora.bianchi@univ-grenoble-alpes.fr}

\author[0000-0001-9249-7082]{Eleonora Bianchi}
\affil{Univ. Grenoble Alpes, CNRS, IPAG, 38000 Grenoble, France}

\author{Ana L\'{o}pez-Sepulcre}
\affil{Univ. Grenoble Alpes, CNRS, IPAG, 38000 Grenoble, France}
\affiliation{Institut de Radioastronomie Millim\'{e}trique, 38406 Saint-Martin d’H\`{e}res, France}

\author{Cecilia Ceccarelli}
\affil{Univ. Grenoble Alpes, CNRS, IPAG, 38000 Grenoble, France}

\author{Claudio Codella}
\affil{INAF, Osservatorio Astrofisico di Arcetri, Largo E. Fermi 5, I-50125, Firenze, Italy}
\affil{Univ. Grenoble Alpes, CNRS, IPAG, 38000 Grenoble, France}

\author{Linda Podio}
\affil{INAF, Osservatorio Astrofisico di Arcetri, Largo E. Fermi 5, I-50125, Firenze, Italy}

\author{Mathilde Bouvier}
\affil{Univ. Grenoble Alpes, CNRS, IPAG, 38000 Grenoble, France}

\author{Joan Enrique-Romero}
\affil{Departament de Qu\'{i}mica, Universitat Aut\`{o}noma de Barcelona, Bellaterra, 08193, Catalonia, Spain}
\affil{Univ. Grenoble Alpes, CNRS, IPAG, 38000 Grenoble, France}

\begin{abstract}
We present ALMA high-angular resolution ($\sim$ 50 au) observations of the Class I binary system SVS13-A. 
We report images of SVS13-A in numerous interstellar complex organic molecules: CH$_{\rm 3}$OH, $^{13}$CH$_{\rm 3}$OH, CH$_{\rm 3}$CHO, CH$_{\rm 3}$OCH$_{\rm 3}$, and NH$_{\rm 2}$CHO. 
Two hot corinos at different velocities are imaged in VLA4A (V$_{sys}$= +7.7 km s$^{-1}$) and VLA4B (V$_{sys}$= +8.5 km s$^{-1}$).
From a non-LTE analysis of methanol lines we derive a gas density of 3 $\times$ 10$^8$ cm$^{-3}$, and gas temperatures of 140 K and 170 K for VLA4A and VLA4B, respectively. 
For the other species the column densities are derived from a LTE analysis.
Formamide, which is the only N-bearing species detected in our observations, is more prominent around VLA4A, while dimethyl ether, methanol and acetaldehyde are associated with both VLA4A and VLA4B. 
We derive in the two hot corinos abundance ratios of $\sim$ 1 for CH$_{\rm 3}$OH, $^{13}$CH$_{\rm 3}$OH, and CH$_{\rm 3}$OCH$_{\rm 3}$, $\sim$ 2 for CH$_{\rm 3}$CHO, and $\sim$ 4 for NH$_{\rm 2}$CHO.
The present dataset supports a chemical segregation between the different species inside the binary system. 
The emerging picture is that of an onion-like structure of the two SVS13-A hot corinos, caused by the different binding energies of the species, also supported by ad hoc quantum chemistry calculations.
In addition, the comparison between molecular and dust maps suggests that the interstellar complex organic molecules emission originates from slow shocks produced by accretion streamers impacting the VLA4A and VLA4B disks and enriching the gas-phase component.


\end{abstract}


\section{Introduction} \label{sec:intro}
The birth of planets inside protostellar disks has been traced back in time, as a result of
recent ALMA high-angular resolution observations showing substructures in the dust distribution of young ($<$1 Myr) disks which may be caused by the interaction of the disk material with forming planets \citep{Sheehan2017, Sheehan2020, Segura-Cox2020}.
Protostellar Class 0/I disks are thus the ideal environment to investigate the initial conditions and the chemical content which will be at least partially inherited by forming planets \citep[e.g.][]{Oberg2021}.
From a chemical point of view, close binaries are unique laboratories as they are expected to originate from the same parent core, i.e. the same initial gas composition and similar gas conditions, in terms of temperature, density and UV illumination. Observed differences in the chemistry of the two binary  components are then expected to be the result of a chemical evolution inside the system \citep[e.g.][]{Manigand2019}.
SVS13-A is a perfect case of study since it is a very well studied Class I protostellar system in NGC1333, hosting a rich chemistry including emission from several interstellar complex organic molecules \citep[hereafter iCOMs; ][]{Lopez2015,Codella2016,Bianchi2017a, Desimone2017, Bianchi2019a, Belloche2020, Yang2021, Diaz2021}. The two components of the binary system (VLA4A and VLA4B) have a separation of $\sim$ 0$\farcs$3 ($\sim$ 90 au at the source distance of 299 pc) and they have been imaged in the continuum using ALMA \citep{Tobin2016,Tobin2018,Diaz2021}. Recently, the presence of two hot corinos in both VLA4A and VLA4B has been reported by \citet{Diaz2021}. The detection of ethylene glycol only towards VLA4A suggests the possibility of a different chemistry at work in the two binaries,
calling for mapping different iCOMs using different excitation transitions.
In this Letter, we present ALMA high-angular resolution observations, imaging the SVS13-A binary system using a large number of lines (17) due to both O-bearing and N-bearing iCOMs, namely methanol (CH$_{\rm 3}$OH and $^{13}$CH$_{\rm 3}$OH), acetaldehyde (CH$_{\rm3}$CHO), dymethyl ether (CH$_{\rm 3}$OCH$_{\rm 3}$) and formamide (NH$_{\rm 2}$CHO), down to planet-formation scales. 

\section{Observations} \label{sec:obs}

The system SVS13-A was observed with ALMA (2018.1.01461.S) in Band 6.
Data were acquired on 2019 September 8 using the C43-6 configuration, with baselines between 43 m and 5.9 km. 
The observations were centered at
$\alpha_{\rm J2000}$ = 03$^{\rm h}$ 29$^{\rm m}$ 3$\fs$8,
$\delta_{\rm J2000}$ = +31$\degr$ 16$\arcmin$ 03$\farcs$8.
The quasar J0510+1800 was used as bandpass and flux calibrator, while J0336+3218 as phase calibrator. 
The absolute flux calibration uncertainty is 20\%. 
Data were calibrated using the standard ALMA calibration pipeline within \textsc{CASA} \citep{McMullin2007}.
Phase self-calibration has been performed using the \textsc{IRAM-GILDAS}\footnote{\url{http://www.iram.fr/IRAMFR/GILDAS}} package, after the determination of line-free continuum channels, and the solutions applied both to the continuum and the spectral cubes.
The observed spectral windows, as well as the synthesized beam and the r.m.s. noise of the continuum-subtracted line-cubes, using spectral channels of 122 kHz (0.15 km s$^{-1}$), are reported in Table \ref{Tab:obs}.

\section{Results} \label{sec:res}

\begin{figure*}[ht]
\centering
\includegraphics[scale=0.7]{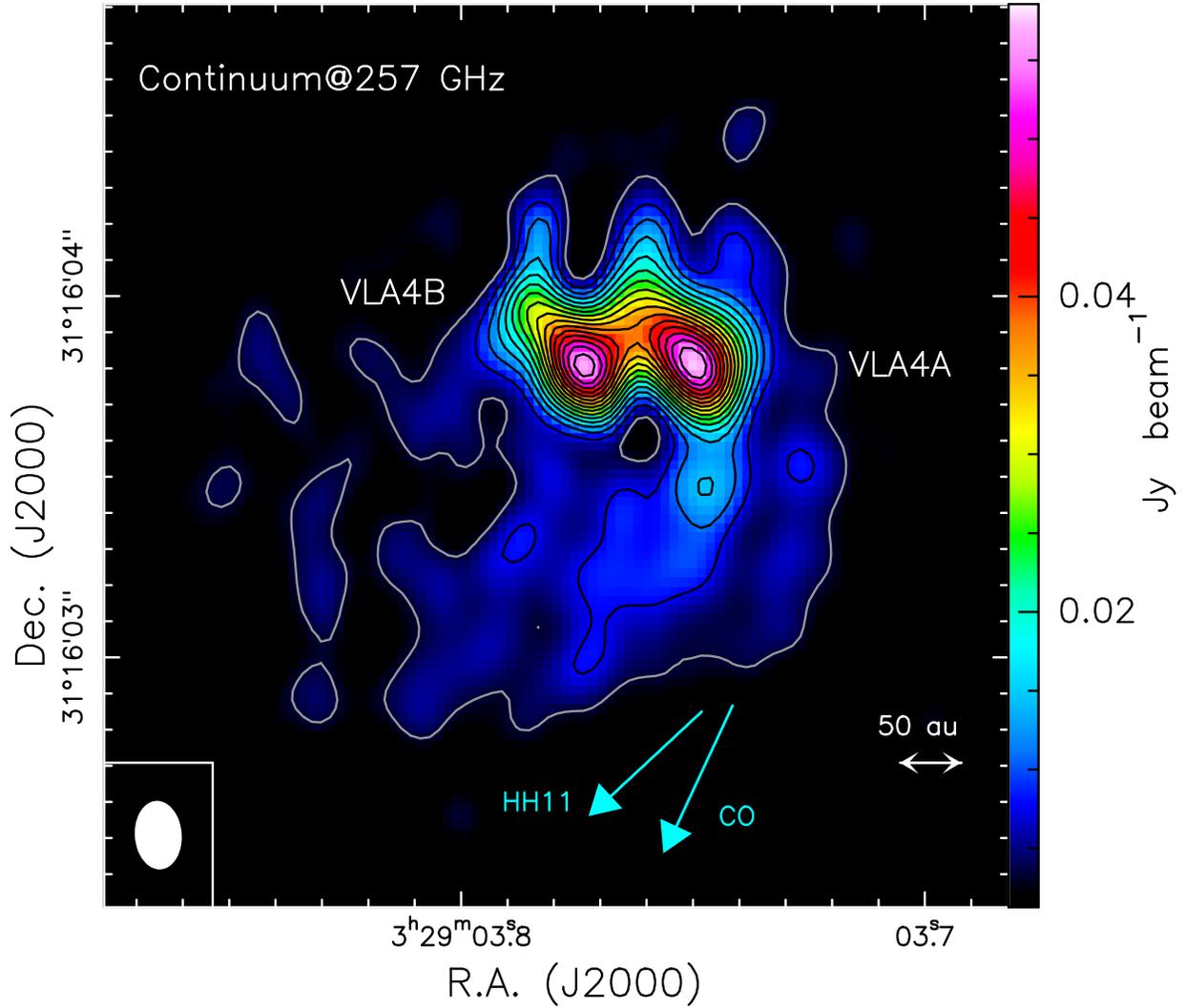}
\caption{Dust continuum emission at 257 GHz as observed by ALMA (colour scale and contours). First contours (gray) and steps (black) are 8$\sigma$, corresponding to 3.35 mJy beam$^{-1}$. The beam size is 0$\farcs$19 $\times$ 0$\farcs$13 (PA=+5$^{\circ}$). The binary system is composed of VLA4A (right) and VLA4B (left). Coordinates are $\alpha_{\rm J2000}$ = 03$^{\rm h}$ 29$^{\rm m}$ 3$\fs$75, $\delta_{\rm J2000}$ = +31$\degr$ 16$\arcmin$ 03$\farcs$81 and $\alpha_{\rm J2000}$ = 03$^{\rm h}$ 29$^{\rm m}$ 3$\fs$773, $\delta_{\rm J2000}$ = +31$\degr$ 16$\arcmin$ 03$\farcs$81 for VLA4A and VLA4B, respectively. 
The blue arrows indicate the directions toward HH11 (133$\degr$) as well as the
main axis of the H$_{\rm 2}$/CO jet (155$\degr$) as inferred by \citet{Lefevre2017}.
\label{fig:continuum}}
\end{figure*}

\subsection{Continuum emission} \label{subsec:cont}

Figure \ref{fig:continuum} shows the SVS13-A map of the dust continuum emission at 1.3mm.
The two emission peaks correspond to the two components of the SVS13-A system previously observed with ALMA as well as at cm-wavelengths \citep{Carrasco2008, Tobin2018, Diaz2021}: VLA4A and VLA4B. 
In addition, the map shows arc-like structures in agreement with what observed by \citet{Tobin2016,Tobin2018} and \citet{Diaz2021}: a bright arc (up to 32$\sigma$) extending from the southern edge of the continuum associated with VLA4A towards the south-east, and two fainter arcs (up to 16$\sigma$) parallel to the first one and extending on the east (i.e. from the southern edge of the continuum associated with VLA4B) and on the west side of VLA4A. 
In addition, the present image shows further three elongated structures extend from the northern edge of the continuum associated with VLA4A and B towards the north, with a slightly negative PA. 
As the ALMA beam is roughly elongated along the N-S direction (PA =+5$\degr$) , we checked for the presence of possible artefacts cause by the cleaning process by repeating the cleaning using  different  tapering.  With  a  larger  circular  beam all the elongated arc-like structures, including the three structures towards the north, are less evident but still present. The presence of accretion streamers, suggested for VLA4B \citep{Diaz2021}, is then confirmed by the present ALMA dataset also towards VLA4A and it has an impact on the following discussion on the origin of iCOMs emission (Sec. \ref{subsec:morphology}).

The continuum emission associated with VLA4A and VLA4B has a disk-like structure. 
We modelled the continuum emission in the \textit{uv} plane 
assuming two elliptical inclined disks centered on the continuum peak positions. 
The best fitting is obtained with one disk of axes 0$\farcs$602 (0$\farcs$001) $\times$ 0$\farcs$231 (0$\farcs$077) and PA 41.1$\degr$ (0.1$\degr$) centered towards VLA4A and a second disk of axes 0$\farcs$325 (0$\farcs$242) $\times$ 0$\farcs$205 (0$\farcs$153) and PA 33.8$\degr$ (0.1$\degr$) centered towards VLA4B. 
The integrated fluxes are 183.15 (0.87) mJy and 126.06 (0.55) mJy for VLA4A and VLA4B, respectively. 
The sizes are: $\sim$ 180 $\times$ 70 au (VLA4A), and 90 $\times$ 70 au (VLA4B). From the axis ratio we can infer an inclination of 67$\degr$ and 51$\degr$ (with 
respect to the plane of the sky) for the VLA4A and VLA4B disks, respectively.
We calculate the disk masses 
assuming isothermal and optically thin dust emission. If we assume a dust temperature of 140 K, as derived from the lines analysis, a dust-to-gas ratio of 1/100, and a dust mass opacity coefficient k$_{\rm \nu}$ of 1.0 cm$^2$ g$^{-1}$ \citep{Ossenkopf1994}, we obtain 0.029 M$_{\odot}$ for VLA4A and 0.020 M$_{\odot}$ for VLA4B. The value obtained for VLA4B is consistent with what reported by \citet{Diaz2021} while the one for VLA4A is higher of a factor 3.
If we assume, a lower dust temperature of 30 K, we obtain 0.159 M$_{\odot}$ for VLA4A and 0.110 M$_{\odot}$ for VLA4B, in good agreement with what found by \citet{Tobin2018}.

\subsection{Emission from iCOMs} \label{subsec:iCOMs}
\begin{figure*}[ht]
\centering
\includegraphics[scale=0.9]{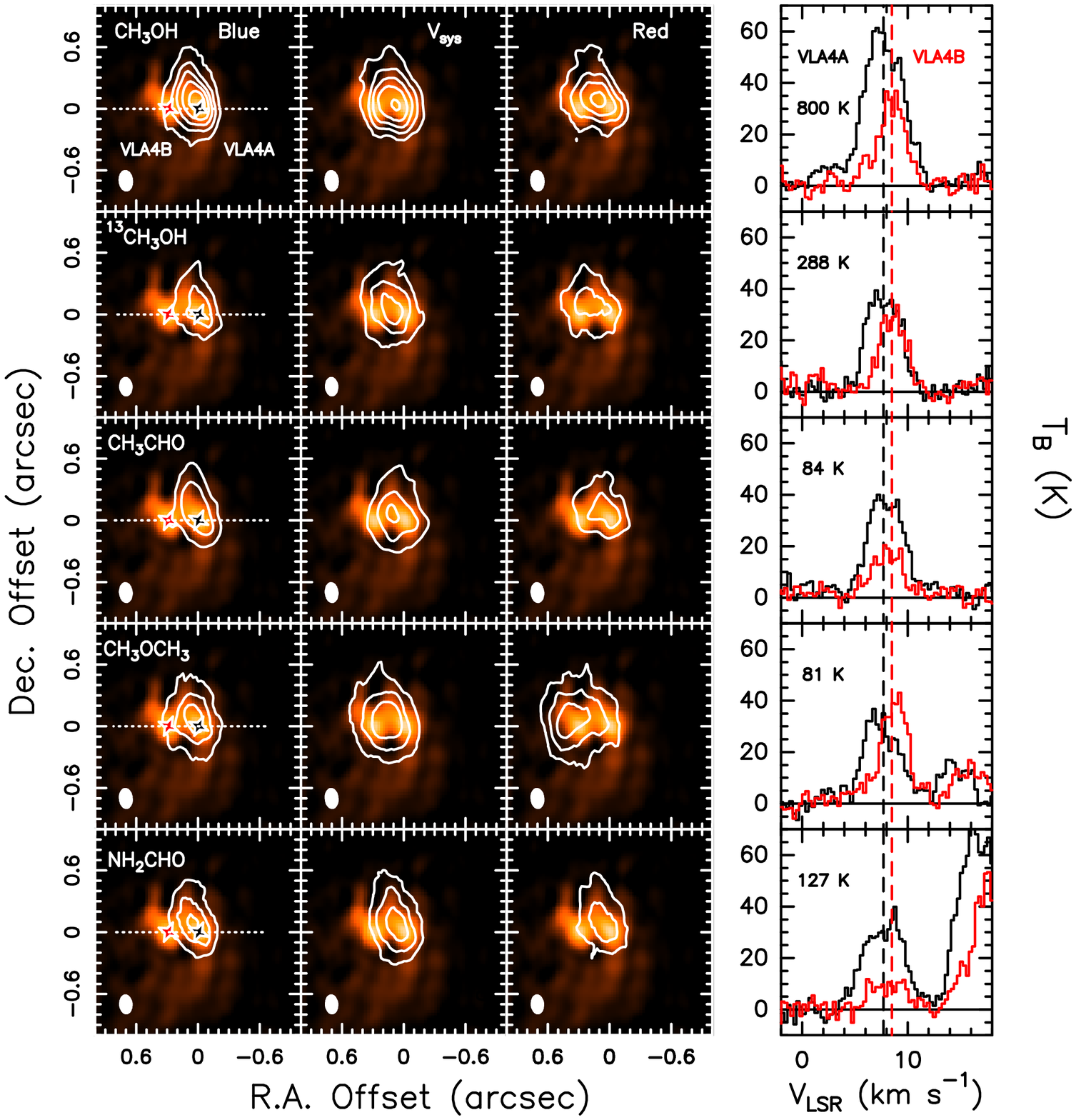}
\caption{Integrated emission from CH$_{\rm 3}$OH, $^{13}$CH$_{\rm 3}$OH, CH$_{\rm 3}$CHO, NH$_{\rm 2}$CHO, and CH$_{\rm 3}$OCH$_{\rm 3}$, in white contours, superposed to the dust emission in colour scale. The emission is integrated in the following velocity intervals: 4 -- 7.5 km s$^{-1}$ in the left panels (V$_{\rm sys}-$), 7.5 -- 8.7 km s$^{-1}$ in the middle panels (V$_{\rm sys}$), and 8.8 -- 12.0 km s$^{-1}$ in the right panels (V$_{\rm sys}+$). First contours and steps are 5$\sigma$, corresponding to 35 mJy beam$^{-1}$, 19 mJy beam$^{-1}$ and 29 mJy beam$^{-1}$, for CH$_{\rm 3}$OH; 34 mJy beam$^{-1}$, 21 mJy beam$^{-1}$ and 28 mJy beam$^{-1}$ for $^{13}$CH$_{\rm 3}$OH; 29 mJy beam$^{-1}$, 19 mJy beam$^{-1}$ and 26 mJy beam$^{-1}$ for CH$_{\rm 3}$CHO; 30 mJy beam$^{-1}$, 18 mJy beam$^{-1}$ and 29 mJy beam$^{-1}$ for CH$_{\rm 3}$OCH$_{\rm 3}$; 33 mJy beam$^{-1}$, 24 mJy beam$^{-1}$ and 27 mJy beam$^{-1}$ for NH$_{\rm 2}$CHO, respectively. The synthesised beams are reported in white in the lower left corner of each panel. The dashed line indicates the axes connecting VLA4A and VLA4B along which the line intensity profiles shown in Fig. \ref{fig:profiles} are extracted. On the right we report the spectra in brightness temperature units, extracted at the continuum peak position of VLA4A (black star) and VLA4B (red star). From $^{13}$CH$_{\rm 3}$OH spectra, we derived a systemic velocity of +7.7 km s$^{-1}$ for VLA4A (black vertical dashed line) and +8.5 km s$^{-1}$ for VLA4B (Red vertical dashed line). Spectra are smoothed at 0.3 km s$^{-1}$ resolution. On the left it is reported the upper level energy of each transition in K.
\label{fig:iCOMs}}
\end{figure*}

We detected a total of 17 emission lines from 4 iCOMs (Table \ref{Tab:lines}). More specifically, we detected 4 lines of CH$_{\rm 3}$OH and 5 lines of $^{13}$CH$_{\rm 3}$OH, covering upper level energies (E$_{\rm up}$) from 57 K to 804 K. We detected also two CH$_{\rm 3}$CHO lines (84 K), 4 lines of CH$_{\rm 3}$OCH$_{\rm 3}$ (26-226 K), and 3 lines of NH$_{\rm 2}$CHO (79-127 K).
In Fig. \ref{fig:iCOMs} we show the emission of one representative line for each  species, integrated on three velocity intervals. Namely, the CH$_{\rm 3}$OH 25$_{\rm 3,22}$--25$_{\rm 2,23}$ A at 241.589 GHz, the $^{13}$CH$_{\rm 3}$OH 14$_{\rm 3,12}$--14$_{\rm 2,13}$ A line at 256.827 GHz, the CH$_{\rm 3}$CHO 13$_{\rm 1,13}$--12$_{\rm 1,12}$ E line at 242.106 GHz, the 13$_{\rm 1,13}$--12$_{\rm 0,12}$ EA, AE, EE, AA transitions at 241.946 GHz and the NH$_{\rm 2}$CHO 12$_{\rm 4,9}$--11$_{\rm 4,8}$ transition at 255.059 GHz. The lines are representative of each species, independently from their upper level energy. All the maps are reported in Figs. \ref{fig:maps1}--\ref{fig:maps5}.
Figure \ref{fig:iCOMs} also shows the iCOMs spectra extracted in the positions corresponding to the continuum peak position of the two sources VLA4A and VLA4B.
From the $^{13}$CH$_{\rm 3}$OH spectra, we reveal different systemic velocities for VLA4A (+7.7 km s$^{-1}$) and VLA4B (+8.5 km s$^{-1}$).
The iCOMs spatial distribution is different for the different species.
For CH$_{\rm 3}$OH and $^{13}$CH$_{\rm 3}$OH, in the interval between +4.0 and +7.5 km s$^{-1}$, the line emission traces VLA4A, with an elongation towards the north. In the velocity intervals 7.5-8.7 km s$^{-1}$, corresponding to the systemic velocity reported for SVS13-A, and 8.8-12.0 km s$^{-1}$ interval, the line emission is associated with both protostars, peaking in between. A beam deconvolved size of 0$\farcs$6 $\times$ 0$\farcs$4 is derived from a simple 2D Gaussian fit in the image plane of the $^{13}$CH$_{\rm 3}$OH moment 0 emission.
For CH$_{\rm 3}$OCH$_{\rm 3}$, the spatial distribution is similar but the emission between +8.8 and +11 km s$^{-1}$ peaks closer to VLA4B. 
Acetaldehyde emission between 8.8 and 11 km s$^{-1}$ peaks closer to VLA4A. Finally, formamide, which is the only N-bearing species detected in our observations, is peaking towards VLA4A.

\begin{figure*}[ht]
\centering
\includegraphics[scale=0.8]{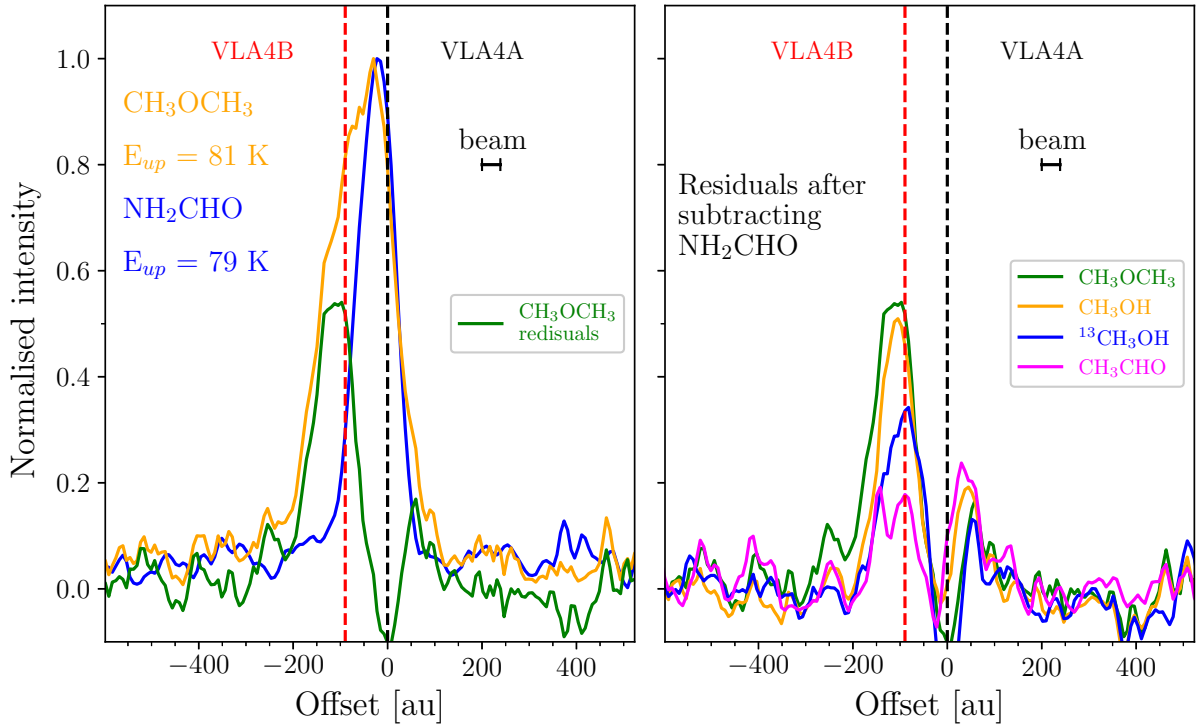}
\caption{\textit{Left Panel:} Normalised intensity profile of the NH$_{\rm 2}$CHO line at 243.521 GHz (blue) and the CH$_{\rm 3}$OCH$_{\rm 3}$ line at 241.946 GHz (orange) extracted along the line connecting VLA4A with VLA4B. The green line shows the redisuals obtained subtracting the formamide emission from the CH$_{\rm 3}$OCH$_{\rm 3}$ profile. The two vertical dashed lines show the position of VLA4A (black) and VLA4B (red). The offset in pixel is given with respect to the position of VLA4A. The pixel size correspond to 25 mas. \textit{Right Panel:} Residuals line profiles of CH$_{\rm 3}$OCH$_{\rm 3}$ (green), CH$_{\rm 3}$OH (orange), $^{13}$CH$_{\rm 3}$OH (blue), and CH$_{\rm 3}$CHO (magenta), after subtracting the formamide emission profile. The analysed lines are the CH$_{\rm 3}$OCH$_{\rm 3}$ line at 241.964 GHz, the CH$_{\rm 3}$OH line at 241.904 GHz, the $^{13}$CH$_{\rm 3}$OH line at 255.204 GHz, the two detected CH$_{\rm 3}$CHO lines at 242.106 GHz and 242.118 GHz, whose intensity is added and, the NH$_{\rm 2}$CHO line at 243.521 GHz.
\label{fig:profiles}}
\end{figure*}

In Table \ref{Tab:lines} we report the integrated line intensities extracted on the continuum peak positions of VLA4A and VLA4B, respectively, and their intensity ratios. The intensities ratios are close to the value of 1 for CH$_{\rm 3}$OCH$_{\rm 3}$, confirming that dimethyl ether is present in both protostars. For methanol and acetaldehyde, intensity ratios vary from 1.3 to 3.2, while formamide abundances are higher in VLA4A by a factor 4 -- 7.
The intensity ratios suggest a chemical differentiation inside the system.
To further investigate the species spatial distribution, 
we compare emission lines with similar E$_{\rm up}$ (between 57 and 81 K) in order to minimise excitation conditions effects.
Figure \ref{fig:profiles} shows the normalised integrated line profiles of CH$_{\rm 3}$OCH$_{\rm 3}$ and NH$_{\rm 2}$CHO, extracted along the horizontal axes connecting the two protostars (Fig. \ref{fig:iCOMs}). The x-axis reports the offset in au with respect to the position of VLA4A. Two vertical dashed lines indicate the positions of the two protostars. The green line is the residual emission obtained subtracting the NH$_{\rm 2}$CHO spectra to the CH$_{\rm 3}$OCH$_{\rm 3}$ emission. If we assume that NH$_{\rm 2}$CHO traces mainly VLA4A, the green profile is the additional emission from VLA4B. Figure \ref{fig:profiles} also shows the residual profiles of CH$_{\rm 3}$OCH$_{\rm 3}$, CH$_{\rm 3}$OH, $^{13}$CH$_{\rm 3}$OH and, CH$_{\rm 3}$CHO after subtracting the NH$_{\rm 2}$CHO emission. The figure highlights the emission of different species at the position of VLA4B. 
Finally, Fig. \ref{fig:MOM1} shows the intensity-weighted peak velocity
(moment 1) maps of the brightest $^{13}$CH$_{\rm 3}$OH, CH$_{\rm 3}$OCHO and NH$_{\rm 2}$CHO lines. All the lines show velocity gradients roughly orthogonal to the outflow direction. Methanol shows the broadest velocity interval while formamide shows the tightest. 

\begin{table*}
\begin{center}
    
	\caption{Properties of 
	the lines detected towards SVS13-A.}
	\label{Tab:lines}
	\begin{tabular}{lccccccc} 
		\hline
Transition & $\nu$$^{\rm a}$ & E$_{\rm up}$$^{\rm a}$ & $S\mu^2$$^{\rm a}$ & \multicolumn{2}{c}{I$_{\rm int}$$^{\rm b}$}& \\
 & (GHz) & (K) & (D$^2$) & \multicolumn{2}{c}{(K km s$^{-1}$)}&\\
\hline
&&&& VLA4A & VLA4B& VLA4A/VLA4B\\
\hline
\vspace{0.25cm}
CH$_{\rm 3}$OH 25$_{\rm 3,22}$--25$_{\rm 2,23}$ A& 241.58876 & 804 & 102 & 251 (53) & 112 (26) & 2.2 (0.7) \\

CH$_{\rm 3}$OH 5$_{\rm 2,3}$--4$_{\rm 2,2}$ E & 241.904147 & 61 & 14 & \multirow{2}{*}{454 (95)} & \multirow{2}{*}{353 (74)} & \multirow{2}{*}{1.3 (0.4)}\\
\vspace{0.25cm}
CH$_{\rm 3}$OH 5$_{\rm -2,4}$--4$_{\rm -2,3}$ E & 241.904643 & 57 & 13\\
CH$_{\rm 3}$OH 16$_{\rm 3,14}$--16$_{\rm 2,15}$ A & 255.241888 & 365 & 59 & 410 (86) & 267 (56) & 1.5 (0.5)\\
\hline
$^{\rm 13}$CH$_{\rm 3}$OH 4$_{\rm 3,2}$--4$_{\rm 2,3}$ A  & 255.203728 & 73 & 3 & 138 (31) & 102 (23) & 1.4 (0.4)&\\ 
$^{\rm 13}$CH$_{\rm 3}$OH 3$_{\rm 3,0}$--3$_{\rm 2,1}$ A & 255.210605 & 64 & 2 & 110 (25) & 76 (18)&\\ 
$^{\rm 13}$CH$_{\rm 3}$OH 7$_{\rm 3,5}$--7$_{\rm 2,6}$  A & 255.214891 & 113 & 6 & 195 (42)  & 148 (32)&\\ 
$^{\rm 13}$CH$_{\rm 3}$OH 3$_{\rm 3,1}$--3$_{\rm 2,2}$ A$^{\rm c}$ & 255.220865 & 64 & 2 & $<$ 118 &  $<$ 80 & \\ 
$^{\rm 13}$CH$_{\rm 3}$OH 14$_{\rm 3,12}$--14$_{\rm 2,13}$ A & 256.826572 & 288 & 13 & 147 (32) &  91 (21)& 1.6 (0.5)\\
\hline
CH$_{\rm 3}$CHO 13$_{\rm 1,13}$--12$_{\rm 1,12}$ E  & 242.106020 & 84 & 163 & 161 (35) & 66 (16) & 2.4 (0.8)\\
CH$_{\rm 3}$CHO 13$_{\rm 1,13}$--12$_{\rm 1,12}$ A  & 242.118136 & 84 & 163 & 155 (35) & 48 (13) & 3.2 (1.1)\\

\hline

CH$_{\rm 3}$OCH$_{\rm 3}$ 5$_{\rm 3,2}$--4$_{\rm 2,3}$ AE  & 241.523808 & \multirow{4}{*}{26} & 28 & \multirow{4}{*}{$>$51$^{\rm d}$} & \multirow{4}{*}{$>$41$^{\rm d}$} & \\
CH$_{\rm 3}$OCH$_{\rm 3}$ 5$_{\rm 3,2}$--4$_{\rm 2,3}$ EA & 241.528318 &  & 15 &  & \\
CH$_{\rm 3}$OCH$_{\rm 3}$ 5$_{\rm 3,2}$--4$_{\rm 2,3}$ EE & 241.528719 &  & 69 &  & \\
\vspace{0.25cm} 
CH$_{\rm 3}$OCH$_{\rm 3}$ 5$_{\rm 3,2}$--4$_{\rm 2,3}$ AA  & 241.531026 &  & 46 &  & \\

CH$_{\rm 3}$OCH$_{\rm 3}$ 21$_{\rm 3,18}$--20$_{\rm 4,17}$ AA  & 241.635773 & \multirow{4}{*}{226} & 61 & \multirow{4}{*}{$<$58 $^{\rm e}$} & \multirow{4}{*}{$<$60  $^{\rm e}$} & \\
CH$_{\rm 3}$OCH$_{\rm 3}$ 21$_{\rm 3,18}$--20$_{\rm 4,17}$ EE & 241.637303 &  & 97 &  & \\
CH$_{\rm 3}$OCH$_{\rm 3}$ 21$_{\rm 3,18}$--20$_{\rm 4,17}$ AE & 241.638831 &  & 37 &  & \\
\vspace{0.25cm} 
CH$_{\rm 3}$OCH$_{\rm 3}$ 21$_{\rm 3,18}$--20$_{\rm 4,17}$ EA  & 241.638834 &  & 24 &  & \\

CH$_{\rm 3}$OCH$_{\rm 3}$ 13$_{\rm 1,13}$--12$_{\rm 0,12}$ EA  & 241.946249 & \multirow{4}{*}{81} & 34 & \multirow{4}{*}{131 (29)} & \multirow{4}{*}{136 (30)} &  \multirow{4}{*}{1.0 (0.3)}\\
CH$_{\rm 3}$OCH$_{\rm 3}$ 13$_{\rm 1,13}$--12$_{\rm 0,12}$ AE  & 241.946249 &  & 69 &  & \\
CH$_{\rm 3}$OCH$_{\rm 3}$ 13$_{\rm 1,13}$--12$_{\rm 0,12}$ EE & 241.946542 &  & 274 &  & \\
\vspace{0.25cm} 
CH$_{\rm 3}$OCH$_{\rm 3}$ 13$_{\rm 1,13}$--12$_{\rm 0,12}$ AA  & 241.946835 &  & 103 &  & \\

CH$_{\rm 3}$OCH$_{\rm 3}$ 19$_{\rm 5,14}$--19$_{\rm 4,15}$ AE  & 256.135096 & \multirow{4}{*}{208} & 104 & \multirow{4}{*}{106 (25)} & \multirow{4}{*}{117 (27)} &  \multirow{4}{*}{0.9 (0.3)}\\
CH$_{\rm 3}$OCH$_{\rm 3}$ 19$_{\rm 5,14}$--19$_{\rm 4,15}$ EA  & 256.135164 &  & 69 &  & \\
CH$_{\rm 3}$OCH$_{\rm 3}$ 19$_{\rm 5,14}$--19$_{\rm 4,15}$ EE  & 256.137180 &  & 278 &  & \\
\vspace{0.25cm} 
CH$_{\rm 3}$OCH$_{\rm 3}$ 19$_{\rm 5,14}$--19$_{\rm 4,15}$ AA  & 256.139230 &  & 174 &  & \\

\hline

NH$_{\rm 2}$CHO 12$_{\rm 1,12}$--11$_{\rm 1,11}$ & 243.521044  & 79 & 156 & 267 (59) & 40 (11) & 7 (2)\\

NH$_{\rm 2}$CHO 12$_{\rm 4,9}$--11$_{\rm 4,8}$ & 255.058533  & 127 & 139 & 160 (37) & 42 (12) & 4 (1) \\

NH$_{\rm 2}$CHO 12$_{\rm 4,8}$--11$_{\rm 4,7}$ &  255.078912 & 127 & 139 & $>$93$^{\rm f}$ & $>$39$^{\rm f}$ \\
\hline

\end{tabular}\\
\end{center}
$^{\rm a}$ Frequencies and spectroscopic parameters have been provided by
\citet[][and references therein]{Xu2008, Xu1997,Endres2009,Kukolich1971} for CH$_{\rm 3}$OH, $^{13}$CH$_{\rm 3}$OH, CH$_{\rm 3}$OCH$_{\rm 3}$, and NH$_{\rm 2}$CHO, respectively, and retrieved from the Cologne Database for Molecular Spectroscopy\footnote{\url{http://www.astro.uni-koeln.de/cdms/}} \citep{Muller2005}. They were provided by \citet{Kleiner1996} and retrieved from the Jet Propulsion Laboratory molecular database\footnote{\url{https://spec.jpl.nasa.gov/}} \citet{pickett1998} for CH$_{\rm 3}$CHO. 
$^{\rm b}$ Errors on the integrated intensity include 20$\%$ of calibration. 
$^{\rm c}$ Contaminated by the NH$_{\rm 2}$CHO 12$_{\rm 3,10}$--11$_{\rm 3,9}$ line at 255.225651 GHz.
$^{\rm d}$ The line profile is truncated as the line is close to the edge of the spectral window.
$^{\rm e}$ Tentatively detected (line peak below 3$\sigma$ detection limit).
$^{\rm f}$ The NH$_{\rm 2}$CHO 12$_{\rm 3,10}$--11$_{\rm 3,9}$ line at 255.225651 GHz is also detected but not considered for the analysis as strongly contaminated by $^{\rm 13}$CH$_{\rm 3}$OH (see $^{\rm c}$).
\end{table*}



\subsection{Column densities and abundances}\label{subsec:colum-dens}

In order to characterise the two hot corinos, we analysed their methanol emission on the dust-peaks (Table \ref{Tab:lines}) via the non-LTE LVG (Large Velocity Gradient) code \texttt{grelvg}, developed by \citet{Ceccarelli2003}.
We used the collisional coefficients with para-H$_2$, computed by \citet{Rabli2010} between 10 and 200 K for the first 256 levels and provided by the BASECOL database \citep{Dubernet2013}. 
We assumed the A-/E- CH$_3$OH ratio equal to 1 \citep{Flower2006} and  $^{12}$C/$^{13}$C = 60 \citep[e.g.][]{Milam2005}.
To compute the line escape probability as a function of the line optical depth we adopted a semi-infinite slab geometry and a linewidth equal to 4 km s$^{-1}$, as measured.
We ran a large grid of models ($\sim 10000$) to cover the $\chi^2$ surface in the parameters space: the total (A- plus E- ) methanol column density N(CH$_{\rm 3}$OH) from $1\times 10^{13}$ to $2\times 10^{18}$ cm$^{-2}$, the H$_{\rm 2}$ density n$_{\rm H2}$ from $1\times 10^{5}$ to $\times 10^{10}$ cm$^{-3}$ and the temperature T from 40 to 200 K.
We fitted the measured $^{13}$CH$_3$OH-A line intensities by comparing them with those predicted by the model, leaving N($^{13}$CH$_{\rm 3}$OH), n$_{\rm H2}$ and T as free parameters.
We obtained a good fit ($\chi^2 = 1.1$) with the following parameters:
N($^{13}$CH$_{\rm 3}$OH) = ($1.0 \pm 0.2$) $\times 10^{17}$ cm$^{-2}$, T = ($140 \pm 40$) K and n$_{H2}$ = $3 \times 10^{8}$ cm$^{-3}$ with a lower limit of $2 \times 10^{7}$ cm$^{-3}$, indicating that the lines are LTE populated.\footnote{Note that the obtained values are in good agreement with those derived by the non-LTE analysis of eighteen $^{13}$CH$_{\rm 3}$OH lines detected by the IRAM-30m ASAI survey \citep{Bianchi2017a}.} 
Line optical depths are estimated using the non-LTE predictions. The predicted opacity of the $^{13}$CH$_3$OH-A line is between 0.1 to 0.4.
Since the $^{12}$CH$_3$OH-E lines are predicted to be very optically thick ($\tau \sim$60), we derive the column density of the main isotopologue from the $^{13}$CH$_{\rm 3}$OH one, assuming the same T of 140 K and we obtain N(CH$_{\rm 3}$OH) =($6 \pm 1$) $\times 10^{18}$ cm$^{-2}$.
We repeated the same analysis also for VLA4B and obtained the following parameters (reduced $\chi^2 = 0.95$):
N(CH$_{\rm 3}$OH) = ($5 \pm 1$) $\times 10^{18}$ cm$^{-2}$, T = ($170 \pm 50$) K and n$_{\rm H2}$ = $3 \times 10^{8}$ cm$^{-3}$ with a lower limit of $2 \times 10^{7}$ cm$^{-3}$, again indicating that the lines are LTE populated.
Please note that the line with E$_{up}$= 800 K cannot be modeled by the non-LTE code because of the lack of collisional coefficients.
Assuming that the line upper level is LTE populated, the predicted intensity is about a factor 2 lower than the observed one\footnote{Note that we used the same beta escape probability $\beta$ as a function of the line opacity $\tau$ used in the non-LTE LVG code, namely $\beta=\frac{1-exp[-3\tau]}{3\tau}$ \citep{deJong1980}, where $\tau$ is estimated to be 0.3.}.
Yet, very likely this high-lying line is subthermally populated, so that the predicted intensity has even to be considered an upper limit.
It has been already found by other authors that high lying (with E$_{up}\geq$ 500 K) methanol lines are most likely due to radiatively populated methanol rather than collisionally populated, as discussed in \citet{Leurini2007}. 

Since the methanol non-LTE analysis showed that both VLA4A and VLA4B hot corinos have large densities, we carried out a LTE rotational diagram analysis for the other iCOMs, assuming a rotational temperature of 140 K and 170 K for VLA4A and VLA4B, respectively,, as derived from the methanol analysis.
The results of the analysis are reported in Table \ref{Table:abundances}. Synthetic LTE spectra generated using \textit{Weeds} \citep{Maret2011} are reported in Figs. \ref{fig:weeds1}--\ref{fig:weeds4}. The column density ratios between VLA4A and VLA4B are consistent with the line intensity ratios reported in Table \ref{Tab:lines}.
In Table \ref{Table:abundances} we also report the column density ratios of the other iCOMs with respect to methanol, which are in agreement with the ASAI analysis \citep{Bianchi2019a}.

\begin{table*}
\caption{Results of the non-LTE LVG and LTE rotational diagram analysis of the iCOMs emission observed towards VLA4A and VLA4B.}
\begin{center}

\begin{tabular}{lccccccc}
  \hline
    \multicolumn{1}{c}{}&\multicolumn{3}{c}{VLA4A}&\multicolumn{3}{c}{VLA4B}& \multicolumn{1}{c}{VLA4A/VLA4B$^a$}\\
    \hline
     \hline

  \multicolumn{1}{c}{Species} &\multicolumn{1}{c}{T$_{\rm rot}$} &\multicolumn{1}{c}{N$_{\rm tot}$}& \multicolumn{1}{c}{X$_{\rm CH_3OH}$$^b$}&\multicolumn{1}{c}{T$_{\rm rot}$} &\multicolumn{1}{c}{N$_{\rm tot}$}& \multicolumn{1}{c}{  X$_{\rm CH_3OH}$$^b$}&  \multicolumn{1}{c}{}\\
  \multicolumn{1}{c}{} & \multicolumn{1}{c}{  (K)}&\multicolumn{1}{c}{(cm$^{-2}$)}& \multicolumn{1}{c}{}& \multicolumn{1}{c}{(K)}&\multicolumn{1}{c}{(cm$^{-2}$)}& \multicolumn{1}{c}{}&  \multicolumn{1}{c}{}\\
\hline
    \multicolumn{8}{c}{non-LTE analysis}\\
\hline
CH$_{\rm 3}$OH$^c$ & 140$^d$ & 6(1) $\times 10^{18}$ & -- & 170$^d$ & 5(1) $\times 10^{18}$ & -- &  \multicolumn{1}{c}{1.2 (0.3)}\\

$^{13}$CH$_3$OH & 140(40) & 1.0(0.2) $\times 10^{17}$ & -- & 170(50) & 8(2) $\times 10^{16}$& --&  \multicolumn{1}{c}{1.3 (0.4)}\\
  \hline
      \multicolumn{8}{c}{Rotational Diagram analysis}\\
\hline
  
  CH$_{\rm 3}$CHO  & 140$^d$ & 2.4(0.4) $\times 10^{16}$ & 4.0 (1.0) $\times 10^{-3}$ &170$^d$ &1.1(0.2) $\times 10^{16}$ & 2.2 (0.6) $\times 10^{-3}$&  \multicolumn{1}{c}{2.2 (0.5)}\\

  CH$_{\rm 3}$OCH$_{\rm 3}$  & 140$^d$ & 7.5(1.2) $\times 10^{16}$ & 1.3 (0.3) $\times 10^{-2}$& 170$^d$ & 1.0(0.2) $\times 10^{17}$ & 2.0 (0.6) $\times 10^{-3}$&  \multicolumn{1}{c}{0.8 (0.2)}\\
  
NH$_{\rm 2}$CHO & 140$^d$ & 5.9(0.9) $\times 10^{15}$ & 4.0 (0.9) $\times 10^{-3}$ & 170$^d$ &1.4(0.3) $\times 10^{15}$ & 2.8 (0.8) $\times 10^{-4}$ &  \multicolumn{1}{c}{4 (1)}\\

 \hline
\end{tabular}\\

$^a$ Column density ratios.
$^b$ Abundance ratio with respect to CH$_{\rm 3}$OH.
$^c$Derived from $^{13}$CH$_{\rm 3}$OH assuming $^{12}$C/$^{13}$C=60.
$^d$ Assumed, as derived by the methanol non-LTE analysis.
 \end{center}
\label{Table:abundances}
\end{table*}

\section{Discussion and Conclusions} \label{sec:disc}
\subsection{Accretion streamers and hot corinos} \label{subsec:morphology}

The three continuum arc-structure observed towards the south and the three-fingers towards the north could be either accretion streamers or alternatively, outflow cavity walls. 
The brightest arc-like structure in the south has been reported on similar spatial scales by \citet{Tobin2018} in the 1.3 mm continuum emission and in blue-shifted C$^{18}$O(2--1) emission.
\citet{Tobin2018} stressed that this one-armed spiral pattern is similar to that
observed in L1448 IRAS3B \citep{Tobin2016} and associated with a
disk inclined by $\sim$ 45$\degr$ surrounding the triple system.
The SVS13-A disks are closer to edge-on (i= 67$\deg$ and 50$\deg$) than L1448 IRAS3B. 
Although the contribution of cavities opened by the precessing
jets/outflows cannot be excluded,
recent studies have shown that accretion streamers feeding disks are a common phenomena from Class 0 to Class II objects \citep{Yen2019,Pineda2020,Garufi2021},
where planet formation is possibly already ongoing. Their presence may strongly affect the disk chemical composition, since they can produce slow shocks when impacting the disk. The chemistry occurring in the shocked gas could synthesise new molecules that would eventually accrete into the disk and, consequently, enrich it.
Further observations of typical shock tracers such as SO and SO$_{\rm 2}$ (e.g. \citealt{Garufi2021}) are needed to confirm the presence of accretion streamers in VLA4A and VLA4B.

On the other hand, iCOMs emission is compact
confirming the presence of two hot corinos, one in VLA4A and a second one in VLA4B. 
The molecular emission sizes (0$\farcs$3--0$\farcs$5) are consistent with thermal sublimation of the icy mantles, namely the classical definition of hot corino \citep{Ceccarelli2004}. 
The velocity gradient roughly perpendicular to the outflow direction (see Fig. \ref{fig:MOM1}) suggests the presence of rotating gas, either from two protostellar disks or from the inner portion of the infalling envelope.

Thanks to the superb angular resolution (0$\farcs$13 in the east-west direction), the blueshifted line emission appears to be elongated in the north direction up to $\sim$150 au, being spatially coincident with the northern accretion streamers (Fig. \ref{fig:maps1}--\ref{fig:maps5}). The iCOMs emission could be associated with material accreting from the streamer and inducing a slow shock in the protostellar disk, similarly to what found in e.g. L1527 \citep{Sakai2014a, Sakai2017, Oya2016}.
Moreover, we cannot exclude the possibility that the system has an higher multiplicity, including multiple unresolved cores.
Observations of different molecular tracers at extremely high-angular resolution (tens of au), as well as an accurate physical model of the system are necessary to conclude about the origin of iCOMs emission.

Finally, the data  suggest a chemical segregation between formamide, tracing mainly VLA4A and the rest of the observed iCOMs, in the inner 100 au of the system.
In Sec. \ref{subsec:opacity} we discuss the possibility that the observed chemical segregation is due to opacity effects. While dust and line opacity effects cannot completely be excluded, our data suggest a real chemical differentiation of the two hot corinos instead of temperature gradients or excitation effect.

\subsection{A real chemical segregation ?} \label{subsec:chem-segr}

A different spatial distribution between O-bearing and N-bearing molecules has been observed in Orion KL \citep[][and references therein]{Blake1987, Peng2013} and in other high-mass star forming regions on scales of thousands of au \citep{Allen2017, Csengeri2019}.
To investigate if a similar segregation is present also in the inner 100 au around low mass protostars is more complicated as only few system have been observed with enough angular resolution.
An intriguing chemical segregation between formamide and acetaldehyde has been observed in the jet-induced shock L1157-B1, at a distance $>$0.1 pc from the protostar \citep{Codella2017}. In this case, the dichotomy is well reproduced by chemical models which assume that the formamide formation is dominated by gas-phase reactions (chemistry effect).
A chemical differentiation has been also observed in the prototypical Class 0 binary system IRAS16293-2422 \citep{Manigand2019}. 
The abundance relative to methanol of CH$_{\rm 3}$OCH$_{\rm 3}$ is similar in the two components of the IRAS16293-2422 system, while it differs by a factor 2 for CH$_{\rm 3}$CHO and 4 for NH$_{\rm 2}$CHO, similarly to what found in VLA4A and VLA4B. In IRAS16293-2422 the observed differences are interpreted as a result of the onion-like structure of the hot corino (physical effect). 
In particular, the NH$_{\rm 2}$CHO rotational temperature is slightly higher (T $\sim$ 140 -- 300 K) than what derived for other iCOMs (T $\sim$ 100 K) \citep{Jorgensen2016,Jorgensen2018, Manigand2019}, suggesting that N-bearing species trace hotter gas, closer to the protostar. On the other hand, recent ALMA observations of other two Class 0 protostars, Perseus B1-c and Serpens S68N, do not show significant difference in the excitations conditions of N-bearing and O-bearing species \citep{Nazari2021, vanGelder2020}, leaving the question open.
Further observations in different N-bearing and O-bearing complex species are required to clarify the origin of the chemical segregation observed in SVS13-A and inferred in IRAS16293-2422 A and B, but not in other low-mass protostars. 
The spatial distribution of iCOMs emission in SVS13-A supports the idea that formamide is formed in a compact region closer to the protostar and thus, contrarily to other iCOMs, can be more easily obscured by the dust. The fact that the dust emission of VLA4B is optically thicker than that of VLA4A supports this scenario.
Note also that the binding energy, recently computed by \citet{Ferrero2020}, of formamide (average $\sim$ 8400 K) is larger than the one of methanol (average $\sim$ 6200 K) in agreement with a possible onion-like structure of the two SVS13-A hot corinos. 
Interestingly, \citet{Diaz2021} report the detection of ethylene glycol only towards VLA4A, suggesting that the observed chemical segregation in SVS13-A is not exclusive of formamide. 
Since no binding energy of ethylene glycol is available in the literature, we carried out new ab initio quantum chemistry calculations to evaluate its binding energy over a 18-water cluster (see Appendix \ref{sec:quant-chem}). 
In order to compare ethylene glycol with formamide, we also recalculated the binding energy of the latter.
We found that the binding energy of ethylene glycol has an average value of $\sim$7100 K, while that of formamide is equal to $\sim$4700 K. We emphasize that the difference with respect to \citet{Ferrero2020} is due to the different used water cluster. The important point here is the relative binding energy of ethylene glycol with respect to formamide, which is not expected to change with the used water cluster. This means that if the calculation by \citet{Ferrero2020} were performed for ethylene glycol, the binding energy would be even higher than that of formamide ($\sim$ 8400 K). This justify the similar spatial distribution of formamide and ethylene glycol, different from that of methanol.
Overall the new calculated binding energies support the hypothesis that the observed chemical differentiation is caused by the onion-like structure of the two SVS13-A hot corinos instead of a spatial segregation between O-bearing and N-bearing species.
Further observations at high-angular resolution (tens of au) with ALMA and at cm-wavelengths with JVLA (and, in perspective, with ngVLA and SKA) are needed to further investigate the chemical stratification inside the two hot corinos VLA4A and VLA4B, and to quantify the dust opacity effects.

\section*{Acknowledgements}
The authors thank the anonymous referee for the constructive comments which substantially helped improving the quality of the paper. This project has received funding from: 1) the European Research Council (ERC) under the European Union's Horizon 2020 research and innovation program, for the Project “The Dawn of Organic Chemistry” (DOC), grant agreement No 741002; 2) the PRIN-INAF 2016 The Cradle of Life - GENESIS-SKA (General Conditions in Early Planetary Systems for the rise of life with SKA); 3) the European Union’s Horizon 2020 research and innovation programs under projects “Astro-Chemistry Origins” (ACO), Grant No 811312.
This paper makes use of the following ALMA data: ADS/JAO.ALMA\#2018.1.01461.S. ALMA is a partnership of ESO (representing its member states), NSF (USA) and NINS (Japan), together with NRC (Canada), MOST and ASIAA (Taiwan), and KASI (Republic of Korea), in cooperation with the Republic of Chile. The Joint ALMA Observatory is operated by ESO, AUI/NRAO and NAOJ.
Most of the computations presented in this paper were performed using the GRICAD infrastructure (https://gricad.univ-grenoble-alpes.fr), which is partly supported by the Equip@Meso project (reference ANR-10-EQPX-29-01) of the programme Investissements d'Avenir supervised by the Agence Nationale pour la Recherche.\\
\textit{Software:} astropy \citep{Astropy1,Astropy2}, matplotlib \citep{Matplotlib}.
\bibliography{Mybib}{}
\bibliographystyle{aasjournal}

\appendix
\counterwithin{figure}{section}
\section{Moment 1 maps}
\begin{figure*}[ht]
\centering

\includegraphics[scale=0.5]{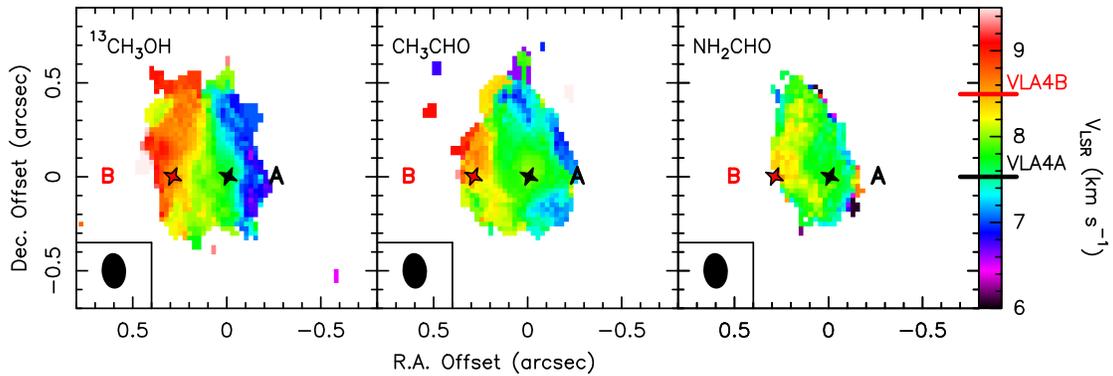}
\caption{Moment 1 (intensity-weighted velocity peak) map of the lines: $^{13}$CH$_{\rm 3}$OH at 256.8265 GHz, CH$_{\rm 3}$CHO at 242.1060 GHz and, NH$_{\rm 2}$CHO at 255.058 GHz (see Tab. \ref{Tab:lines}). The two stars indicate the VLA4A and VLA4B positions. 
The systemic velocities of the two protostars are indicated as
horizontal lines in the wedge.
\label{fig:MOM1}}
\end{figure*}

Figure \ref{fig:MOM1} reports the intensity-weighted velocity peak (moment 1) maps of the following lines: $^{13}$CH$_{\rm 3}$OH at 256.8265 GHz, CH$_{\rm 3}$CHO at 242.1060 GHz and, NH$_{\rm 2}$CHO at 255.058 GHz. The two stars indicate the VLA4A and VLA4B positions.  The systemic velocities of the two protostars are indicated as
horizontal lines in the wedge.

\section{ALMA observations}
Table \ref{Tab:obs} reports the parameters of the observed ALMA setup.
\counterwithin{figure}{section}
\counterwithin{table}{section}

\begin{table*}
\begin{center}
	\caption{Observations}
	\label{Tab:obs}
	\begin{tabular}{lccl} 
		\hline
Frequencies & Synthetised beam & rms & Detected transitions\\
(GHz) & ($\arcsec$) & (mJy/beam kms$^{-1}$ ) & \\
\hline
241.528 -- 241.647 & 0$\farcs$21$\times$0$\farcs$14 & 4.8 & CH$_{\rm 3}$OH 25$_{\rm 3,22}$--25$_{\rm 2,23}$ A\\
 &  &  & CH$_{\rm 3}$OCH$_{\rm 3}$ 5$_{\rm 3,2}$--4$_{\rm 2,3}$ EA, AE, EE, AA\\
 &  &  & CH$_{\rm 3}$OCH$_{\rm 3}$ 21$_{\rm 3,18}$--20$_{\rm 4,17}$ EA,  AE, EE, AA\\
 \hline
 241.900 -- 242.020 & 0$\farcs$21$\times$0$\farcs$14 & 4.9 & CH$_{\rm 3}$OH 5$_{\rm 2,3}$--4$_{\rm 2,2}$ E\\
 &  &  & CH$_{\rm 3}$OH 5$_{\rm -2,4}$--4$_{\rm -2,3}$ E\\
 &  &  & CH$_{\rm 3}$OCH$_{\rm 3}$ 13$_{\rm 1,13}$--12$_{\rm 0,12}$ EA, AE, EE, AA\\
 \hline
 242.076 -- 242.135  &0$\farcs$20$\times$0$\farcs$13 & 4.8 & CH$_{\rm 3}$CHO 13$_{\rm 1,13}$--12$_{\rm 1,12}$ E, A\\
\hline
 243.491 -- 243.550 & 0$\farcs$22$\times$0$\farcs$14 & 6.2 & NH$_{\rm 2}$CHO 12$_{\rm 1,12}$--11$_{\rm 1,11}$\\
 \hline
 255.020 -- 255.080 & 0$\farcs$20$\times$0$\farcs$13 & 5.3 & NH$_{\rm 2}$CHO 12$_{\rm 4,9}$--11$_{\rm 4,8}$\\ 
& & & NH$_{\rm 2}$CHO 12$_{\rm 4,8}$--11$_{\rm 4,7}$\\
 \hline
255.196 -- 255.255  & 0$\farcs$20$\times$0$\farcs$13 & 5.3 & CH$_{\rm 3}$OH 16$_{\rm 3,14}$--16$_{\rm 2,15}$ A\\
 & & & $^{\rm 13}$CH$_{\rm 3}$OH 4$_{\rm 3,2}$--4$_{\rm 2,3}$ A\\
  & & &      $^{\rm 13}$CH$_{\rm 3}$OH 3$_{\rm 3,0}$--3$_{\rm 2,1}$ A\\
  & & &  $^{\rm 13}$CH$_{\rm 3}$OH 7$_{\rm 3,5}$--7$_{\rm 2,6}$  A\\
   & & & $^{\rm 13}$CH$_{\rm 3}$OH 3$_{\rm 3,1}$--3$_{\rm 2,2}$ A\\
\hline
256.107 -- 256.167 & 0$\farcs$19$\times$0$\farcs$13 & 5.1 & CH$_{\rm 3}$OCH$_{\rm 3}$ 19$_{\rm 5,14}$--19$_{\rm 4,15}$ AE, EA, EE, AA\\
\hline
256.797 -- 256.856  & 0$\farcs$19$\times$0$\farcs$13 & 5.6 &  $^{\rm 13}$CH$_{\rm 3}$OH 14$_{\rm 3,12}$--14$_{\rm 2,13}$ A\\
  & & & CH$_{\rm 3}$OCH$_{\rm 3}$ 19$_{\rm 5,14}$--19$_{\rm 4,15}$ AE, EA, EE, AA\\
\hline 
\end{tabular}
\end{center}
\end{table*}
\newpage
\section{Channel maps of the present ALMA dataset}
\counterwithin{figure}{section}
\counterwithin{table}{section}

In this section, we report the images of all the 17 transitions of the 4 iCOMs (CH$_{\rm 3}$OH, $^{13}$CH$_{\rm 3}$OH, CH$_{\rm 3}$CHO, NH$_{\rm 2}$CHO, and CH$_{\rm 3}$OCH$_{\rm 3}$) observed in SVS13-A. \\

\begin{figure*}[ht!]
\centering
\includegraphics[scale=1]{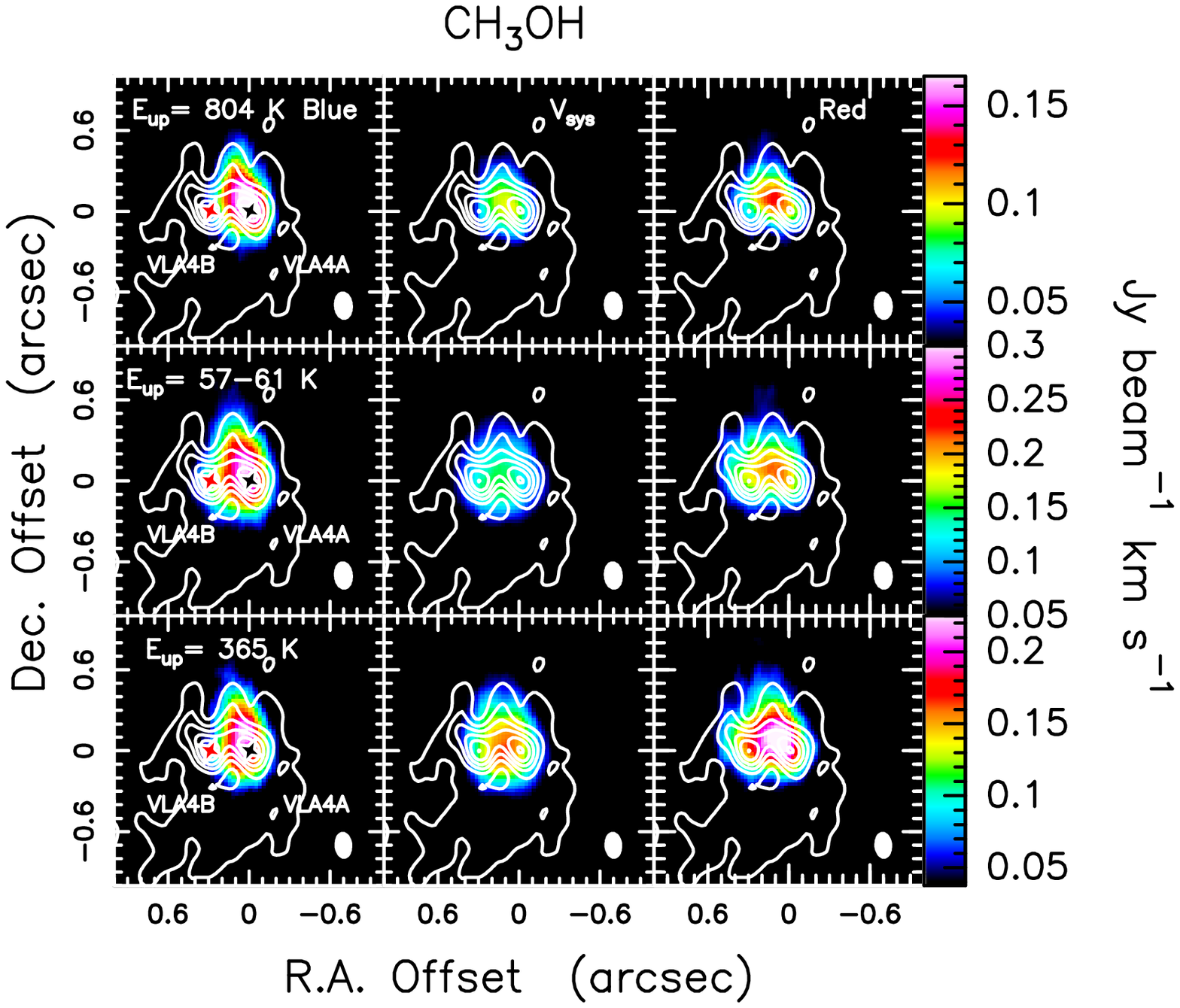}
\caption{Integrated emission of all the detected lines of CH$_{\rm 3}$OH, in colour scale, superposed to the dust emission in white contours. The emission is integrated in the following velocity intervals: 4 -- 7.5 km s$^{-1}$ in the left panels (Blue), 7.5 -- 8.7 km s$^{-1}$ in the middle panels (V$_{\rm sys}$), and 8.8 -- 12.0 km s$^{-1}$ in the right panels (Red). First contours are 10$\sigma$, corresponding to 4 mJy beam$^{-1}$, and steps are 25$\sigma$, corresponding to 10.46 mJy beam$^{-1}$. The synthesised beams are reported in white in the lower right corner of each panel. On the left it is reported the upper level energy of each transition in K.}
\label{fig:maps1}
\end{figure*}

\begin{figure*}[ht!]
\centering
\includegraphics[scale=1]{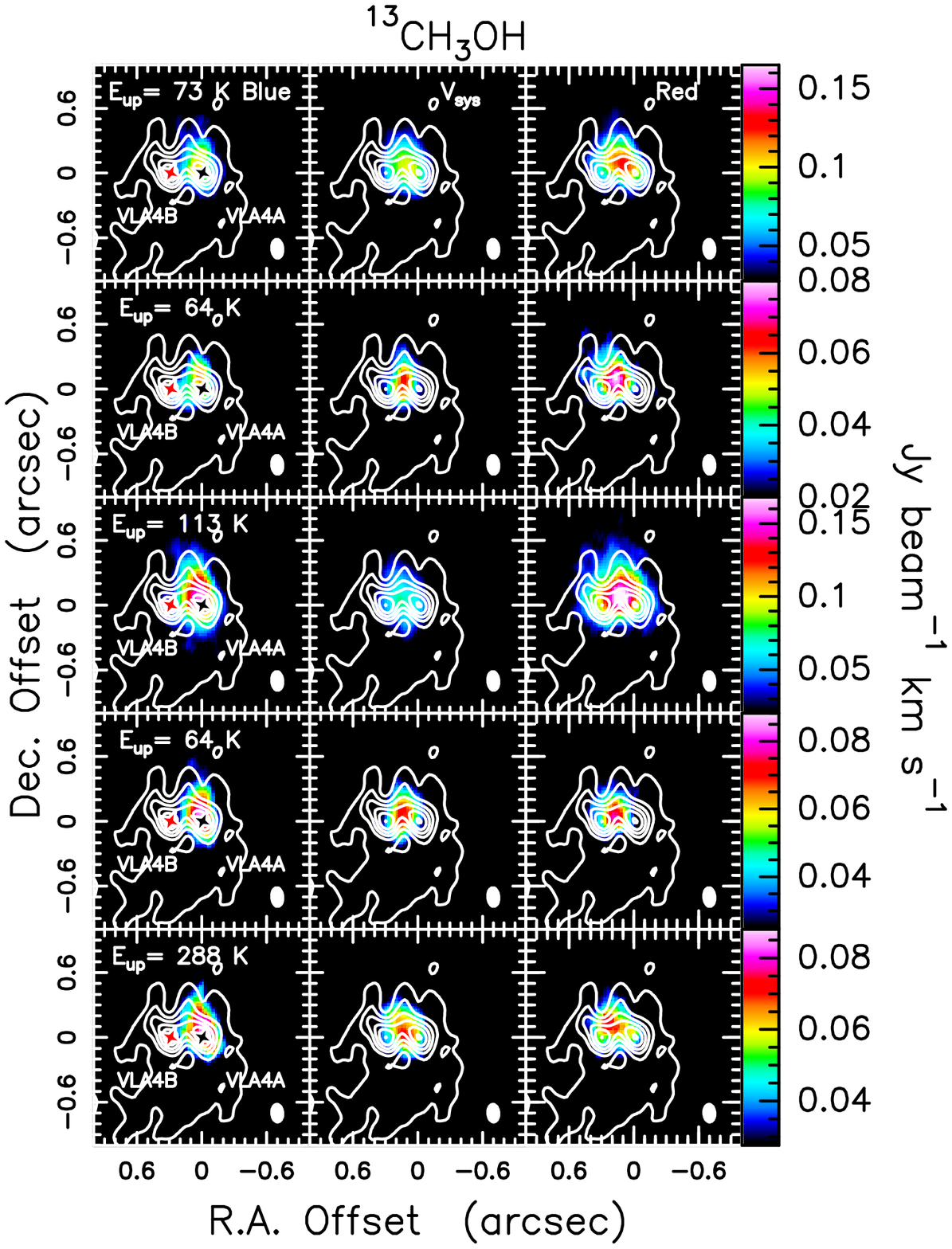}
\caption{Integrated emission of all the detected lines of  $^{13}$CH$_{\rm 3}$OH, in colour scale, superposed to the dust emission in white contours. The emission is integrated in the following velocity intervals: 4 -- 7.5 km s$^{-1}$ in the left panels (Blue), 7.5 -- 8.7 km s$^{-1}$ in the middle panels (V$_{\rm sys}$), and 8.8 -- 12.0 km s$^{-1}$ in the right panels (Red). First contours are 10$\sigma$, corresponding to 4 mJy beam$^{-1}$, and steps are 25$\sigma$, corresponding to 10.46 mJy beam$^{-1}$. The synthesised beams are reported in white in the lower right corner of each panel. On the left it is reported the upper level energy of each transition in K.}
\label{fig:maps2}
\end{figure*}

\begin{figure*}[ht!]
\centering
\includegraphics[scale=1]{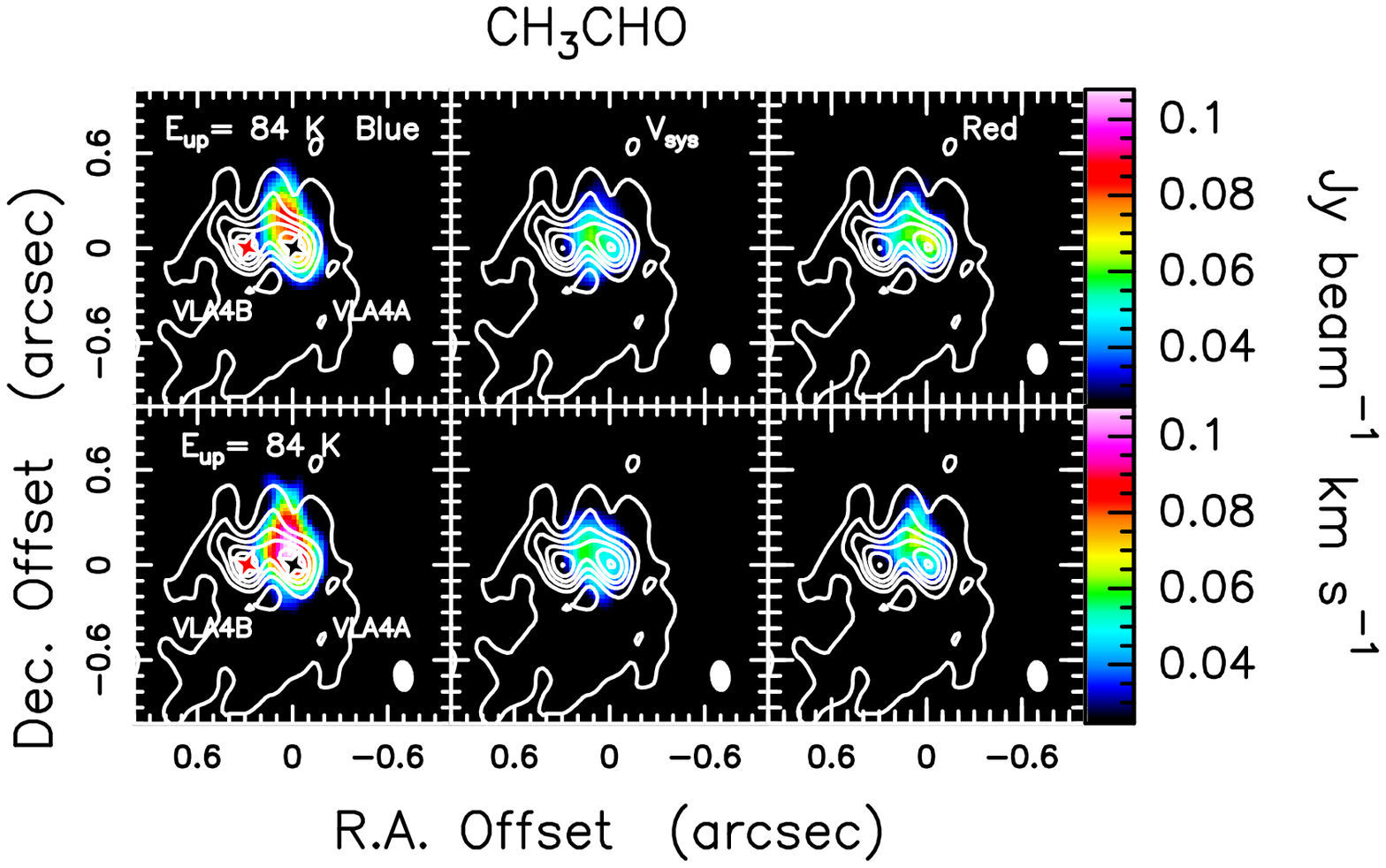}
\caption{Integrated emission of all the detected lines of CH$_{\rm 3}$CHO, in colour scale, superposed to the dust emission in white contours. The emission is integrated in the following velocity intervals: 4 -- 7.5 km s$^{-1}$ in the left panels (Blue), 7.5 -- 8.7 km s$^{-1}$ in the middle panels (V$_{\rm sys}$), and 8.8 -- 12.0 km s$^{-1}$ in the right panels (Red). First contours are 10$\sigma$, corresponding to 4 mJy beam$^{-1}$, and steps are 25$\sigma$, corresponding to 10.46 mJy beam$^{-1}$. The synthesised beams are reported in white in the lower right corner of each panel. On the left it is reported the upper level energy of each transition in K.}
\label{fig:maps3}
\end{figure*}

\begin{figure*}[ht!]
\centering
\includegraphics[scale=1]{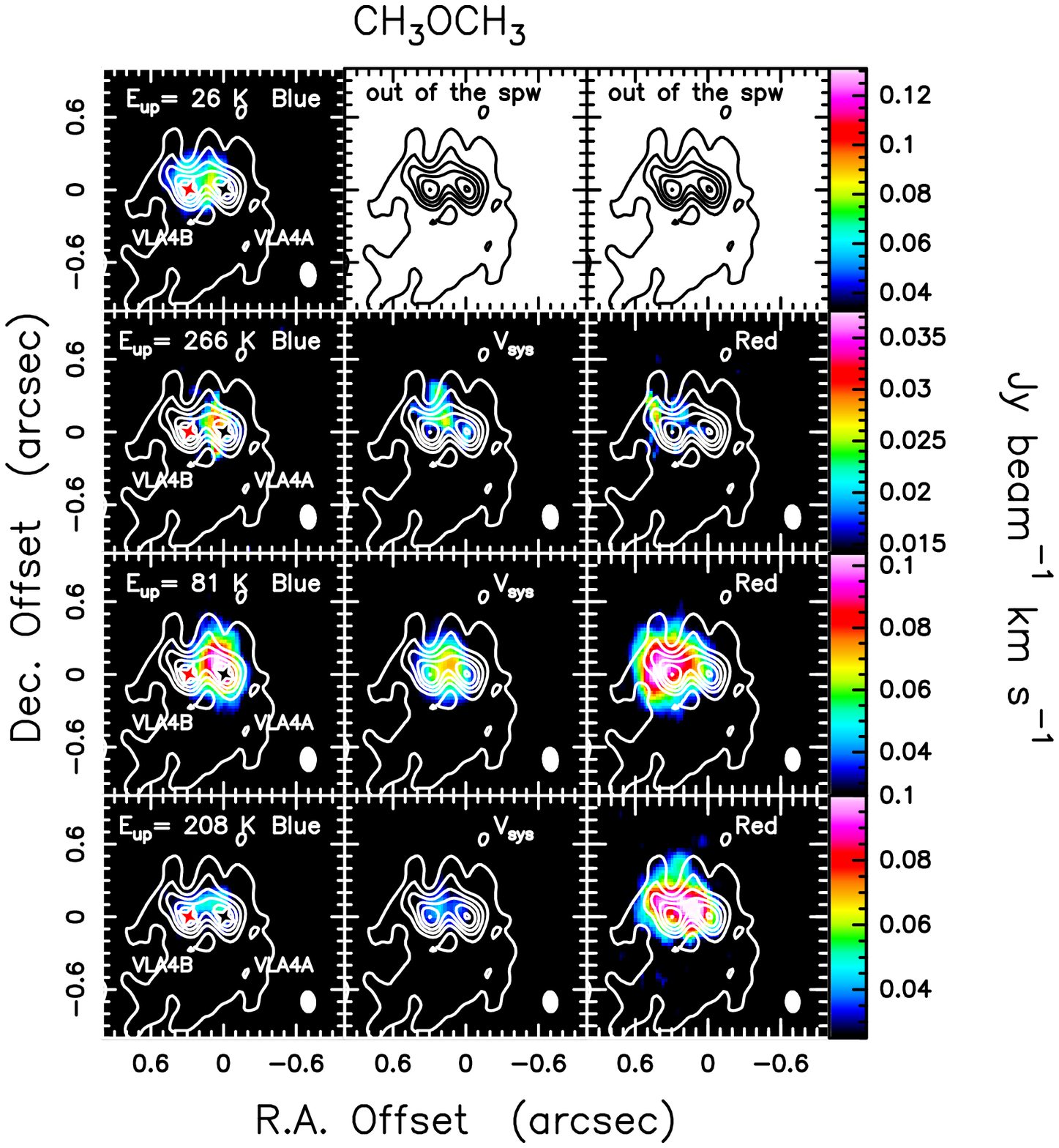}
\caption{Integrated emission of all the detected lines of CH$_{\rm 3}$OCH$_{\rm 3}$, in colour scale, superposed to the dust emission in white contours. The emission is integrated in the following velocity intervals: 4 -- 7.5 km s$^{-1}$ in the left panels (Blue), 7.5 -- 8.7 km s$^{-1}$ in the middle panels (V$_{\rm sys}$), and 8.8 -- 12.0 km s$^{-1}$ in the right panels (Red). First contours are 10$\sigma$, corresponding to 4 mJy beam$^{-1}$, and steps are 25$\sigma$, corresponding to 10.46 mJy beam$^{-1}$. The synthesised beams are reported in white in the lower right corner of each panel. On the left it is reported the upper level energy of each transition in K. The panels showing only the continuum contours refer to line emission partially  outside the observed spectral window (spw). Please note that each line is composed of 4 blended transitions very close in frequency (AE, EA, EE, AE). This structure, due to the rotation in the CH$_{\rm 3}$OCH$_{\rm 3}$ molecule of two functional groups CH$_{\rm 3}$-, leads to some contamination in the channel maps. However, we would like to stress that this velocity-blending does not affect the line spatial distribution.}
\label{fig:maps4}
\end{figure*}

\begin{figure*}[ht!]
\centering
\includegraphics[scale=0.8]{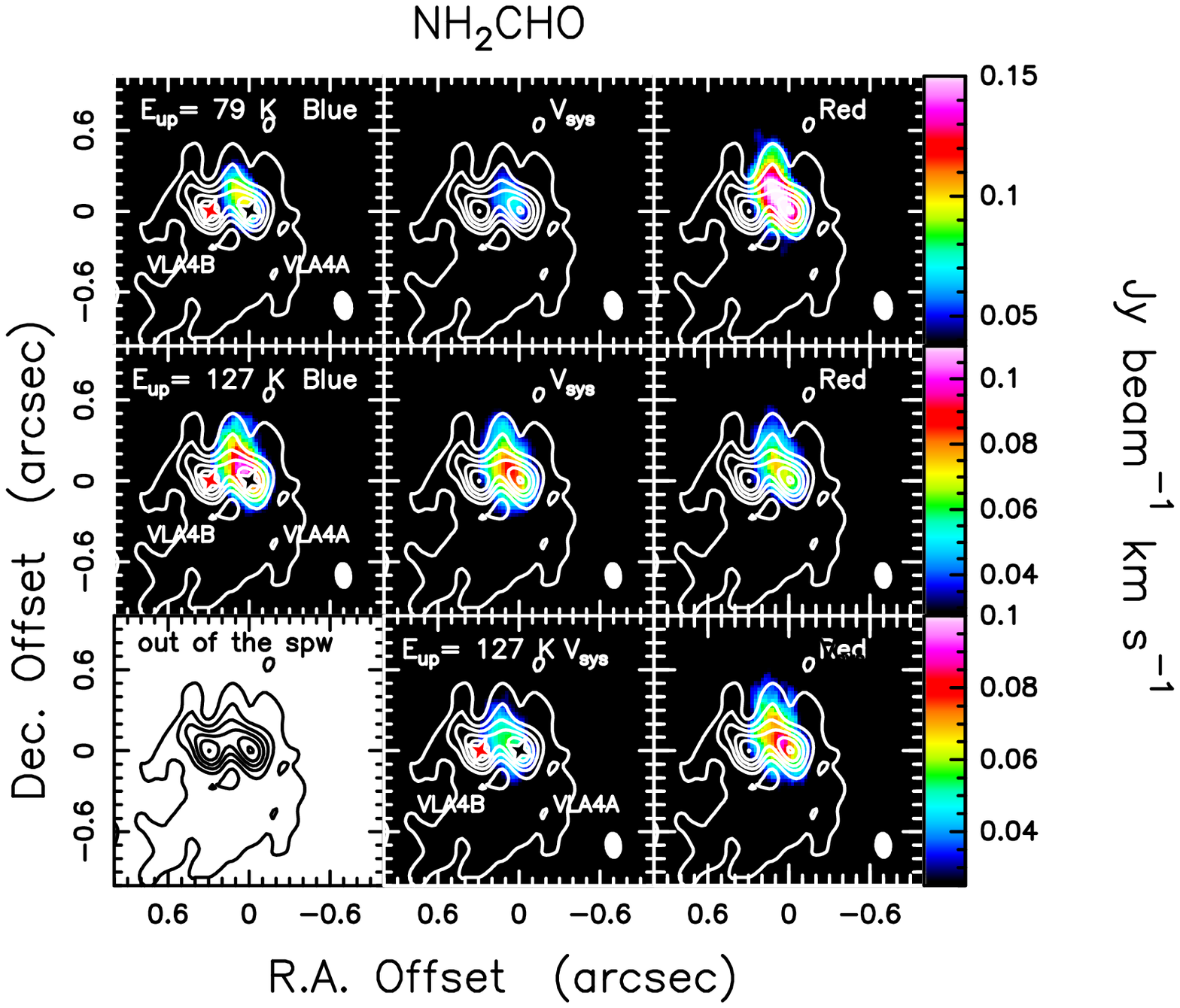}
\caption{Integrated emission of all the detected lines of NH$_{\rm 2}$CHO, in colour scale, superposed to the dust emission in white contours. The emission is integrated in the following velocity intervals: 4 -- 7.5 km s$^{-1}$ in the left panels (Blue), 7.5 -- 8.7 km s$^{-1}$ in the middle panels (V$_{\rm sys}$), and 8.8 -- 12.0 km s$^{-1}$ in the right panels (Red). First contours are 10$\sigma$, corresponding to 4 mJy beam$^{-1}$, and steps are 25$\sigma$, corresponding to 10.46 mJy beam$^{-1}$. The synthesised beams are reported in white in the lower right corner of each panel. On the left it is reported the upper level energy of each transition in K. The panel showing only the continuum contours refers to line emission partially  outside the observed spectral window (spw).}
\label{fig:maps5}
\end{figure*}


\section{Dust and line opacity effects} \label{subsec:opacity}

Previous studies of the binary system IRAS 4A have shown that dust could dramatically obscure molecular emission at (sub-)mm wavelengths \citep{Desimone2020}. In particular, while iCOMs emission is observed only  towards one of the two protostars of the binary system at (sub-)mm wavelengths \citep{Lopez2017}, a similar emission is instead observed at radio wavelengths, less affected by dust opacity \citep{Desimone2020}.
In the case of the binary system SVS13-A, dust opacity effects may also play a role.
In particular, at 1.2 mm the flux density of VLA4A is higher by a factor of 1.4 with respect to that of VLA4B 
while VLA4B is brighter at shorter wavelenghts \citep{Diaz2021}.
Moreover, the spectral index is different towards the two sources (2.2-2.4 in VLA4B and 3.0-3.1 in VLA4A). This suggests a) possible grain growth towards VLA4B implying a large fraction of large dust grains which enhance the flux at 9mm wavelengths or, alternatively, b) optically thick dust at 1.3mm towards VLA4B. In this second hypothesis dust opacity could prevent us from detecting molecular emission towards VLA4B, similarly to the case of IRAS 4A. In any case, as all the observed transitions are at  similar frequencies, dust opacity is expected to affect the emission from iCOMs in the same way if they originate from the same region. 
Another explanation for a different spatial distribution of molecular species could be line opacity. Optically thick lines could trace an external shell of the protostellar envelope, more extended than the one traced by optically thin lines.
The simultaneous observations of lines covering a broad range of upper level energies (26 K - 804 K) allow us to verify that molecular lines from the same species show a similar spatial distribution, regardless of the upper level energy of the transition (see Figs. \ref{fig:maps1}--\ref{fig:maps5}. 
We conclude that the observed iCOMs spatial distribution is not driven by excitation effects.

\section{Comparison with LTE modelling}
We report here the synthetic LTE spectra generated with \textit{Weeds} using the gas parameter derived from the LVG and rotational diagram analysis (see Sec. \ref{subsec:colum-dens}).


\begin{figure*}[ht]
\centering
\includegraphics[scale=0.7]{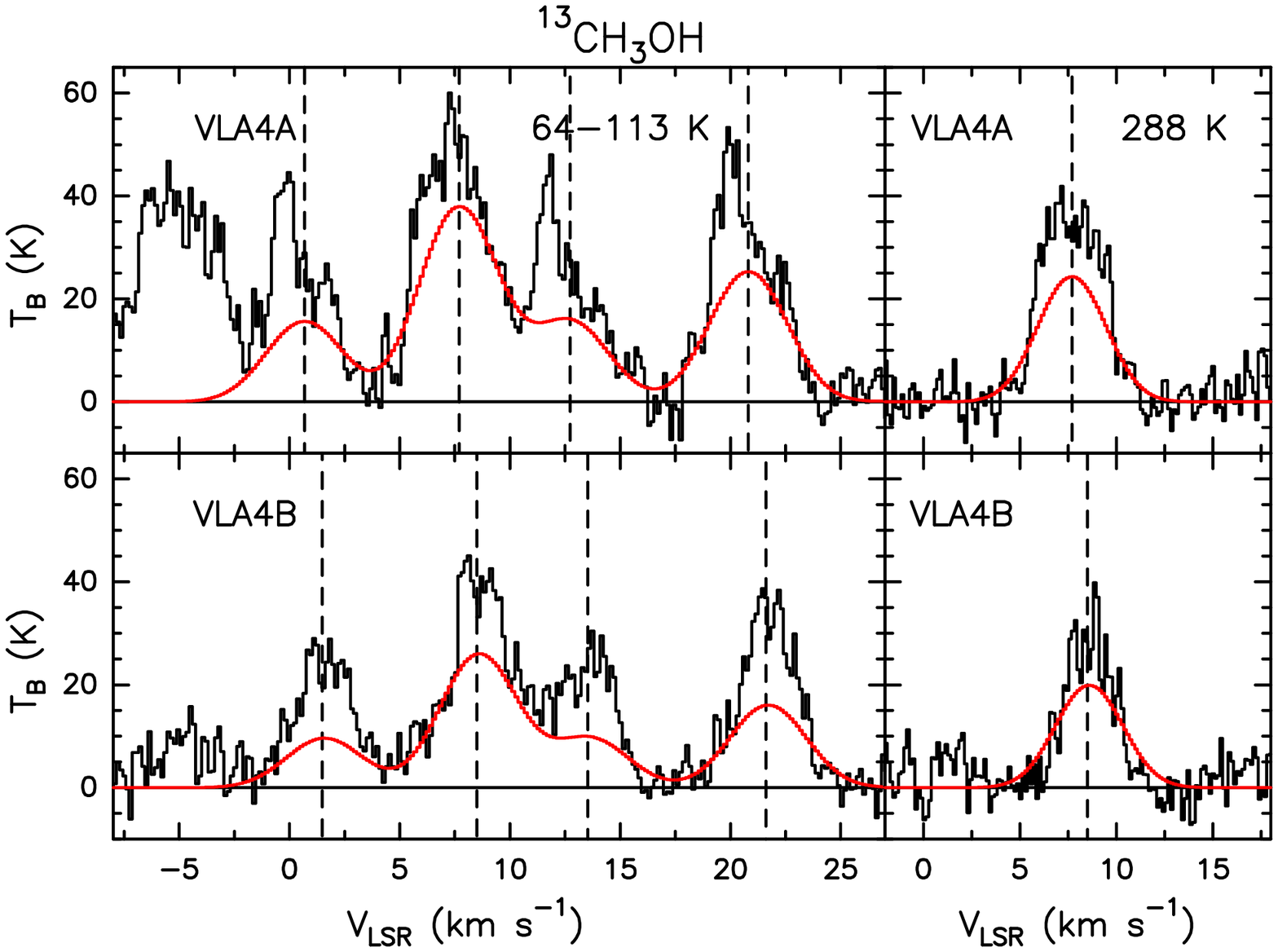}
\caption{Synthetic LTE spectra generated with \textit{Weeds} using the gas parameter in Table \ref{Table:abundances}.}
\label{fig:weeds1}
\end{figure*}

\begin{figure*}[ht]
\centering
\includegraphics[scale=0.7]{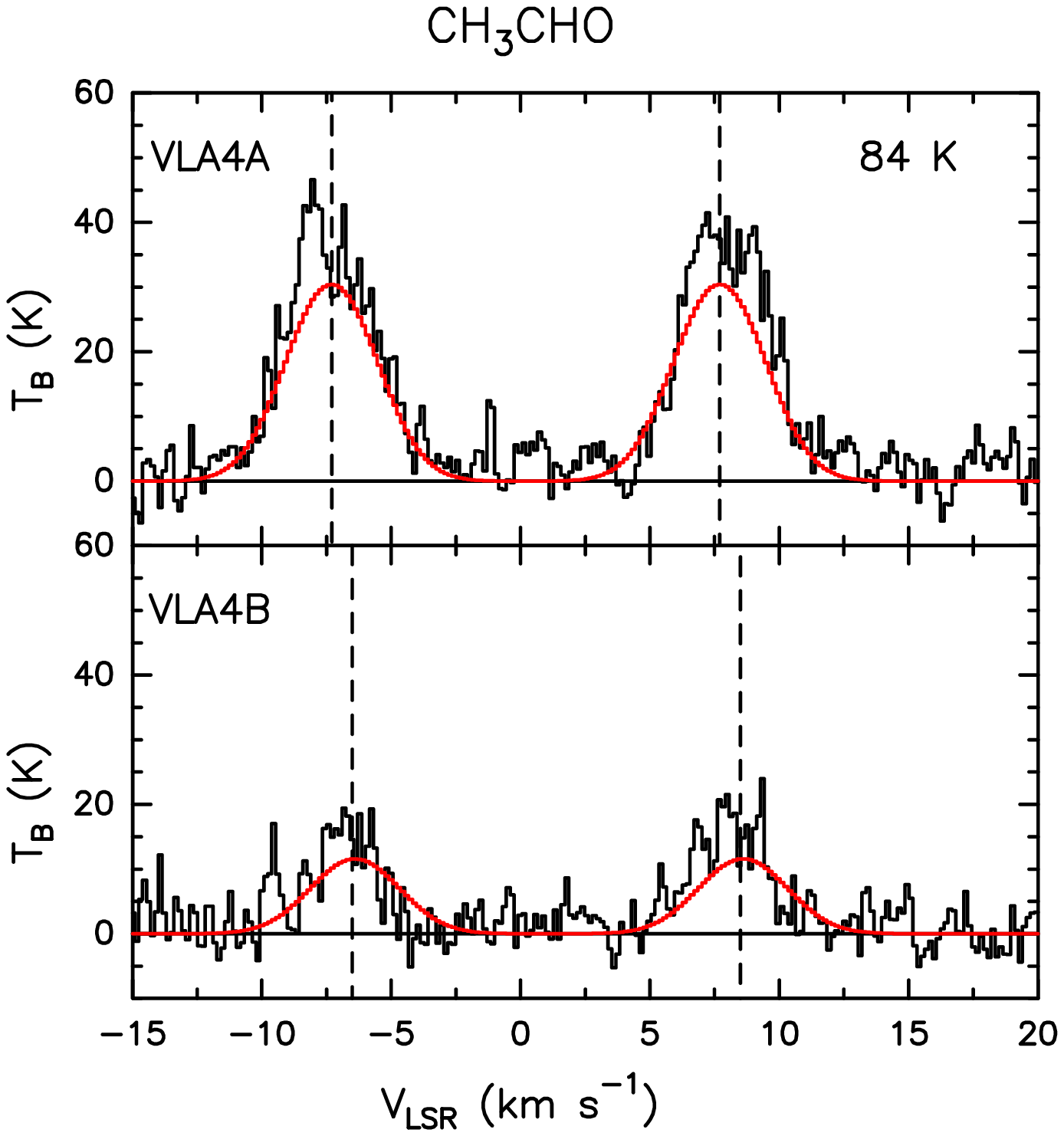}
\caption{Synthetic LTE spectra generated with \textit{Weeds} using the gas parameter in Table \ref{Table:abundances}.}
\label{fig:weeds2}
\end{figure*}

\begin{figure*}[ht]
\centering
\includegraphics[scale=0.7]{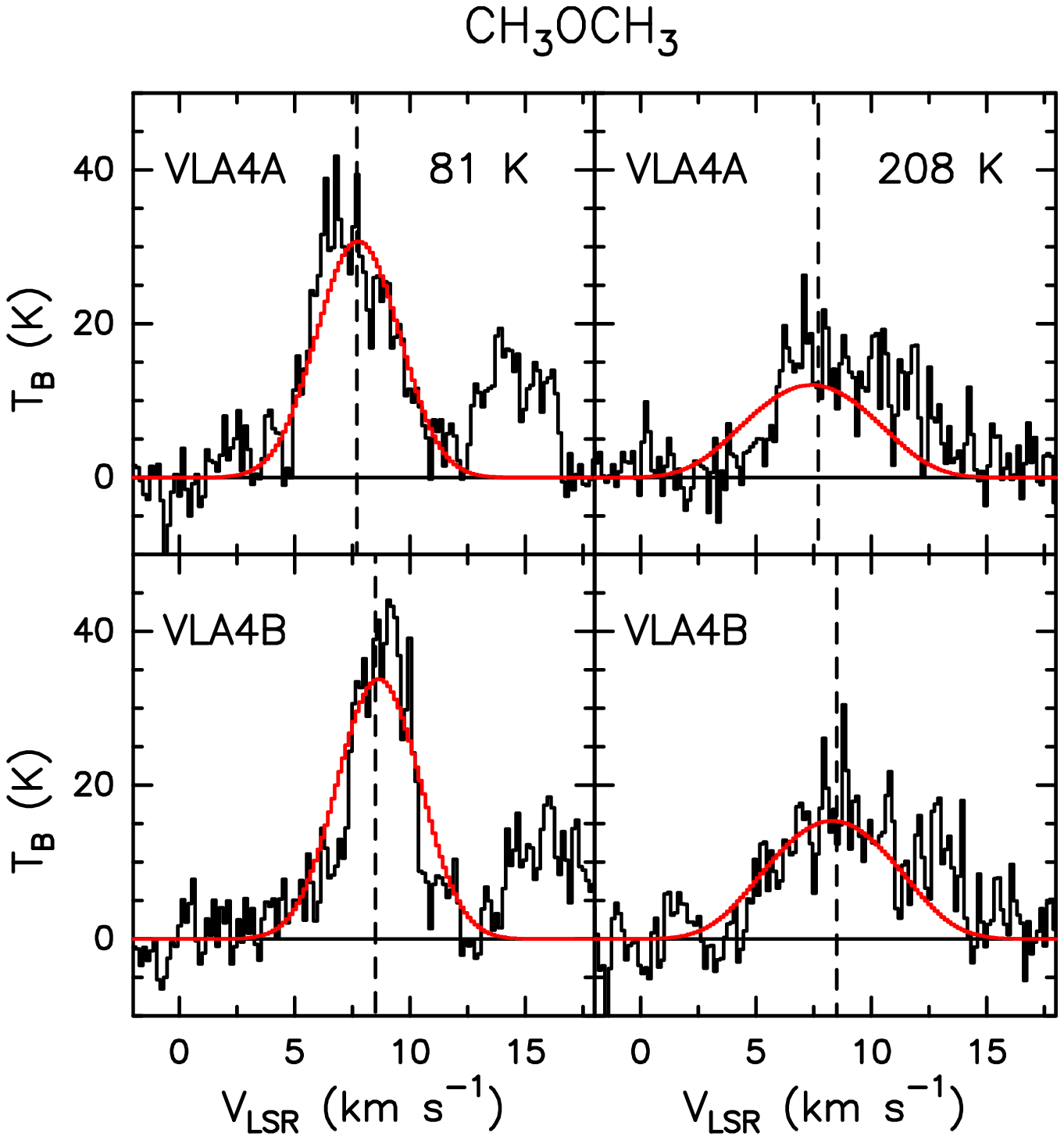}
\caption{Synthetic LTE spectra generated with \textit{Weeds} using the gas parameter in Table \ref{Table:abundances}.}
\label{fig:weeds3}
\end{figure*}

\begin{figure*}[ht]
\centering
\includegraphics[scale=0.7]{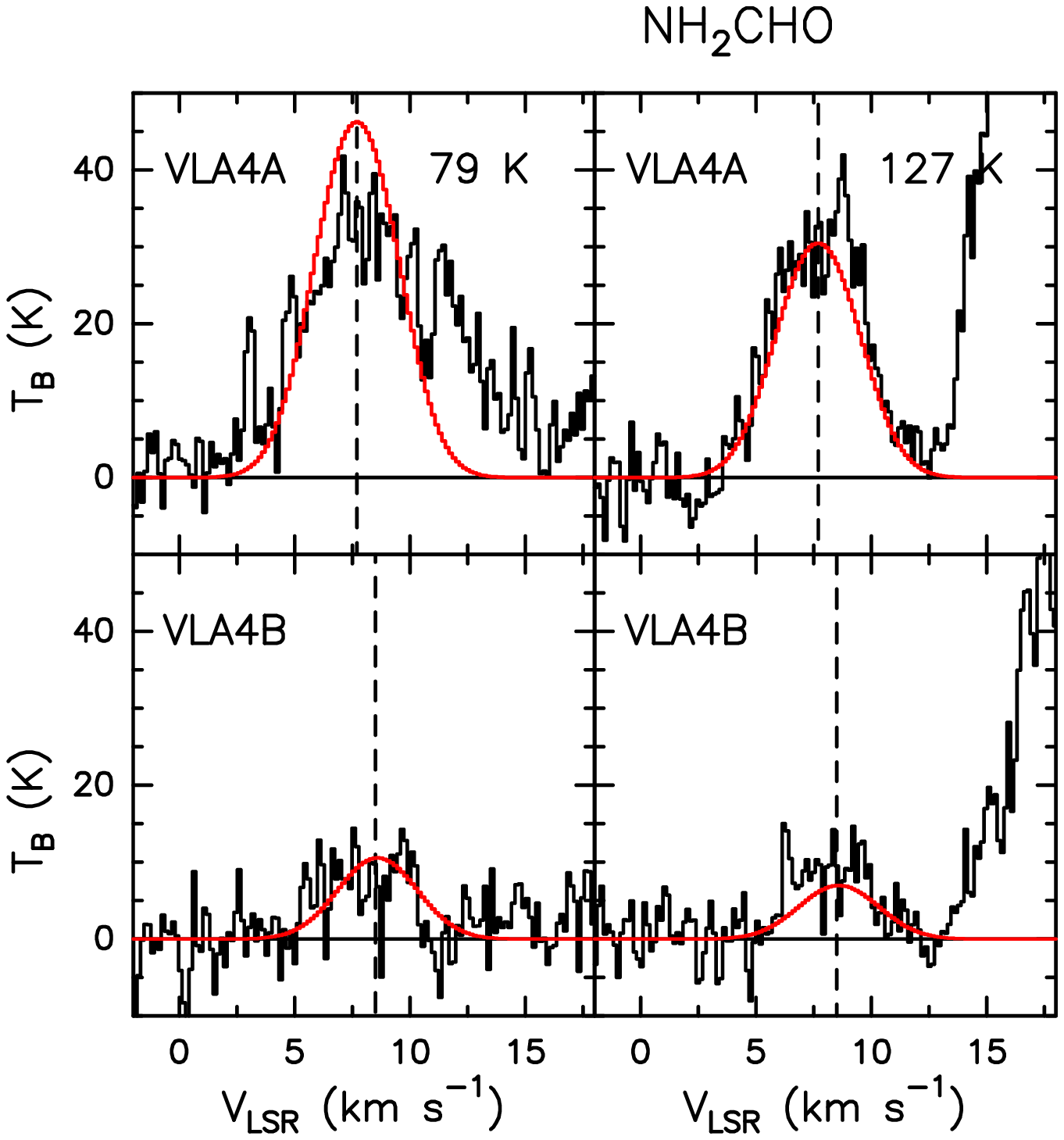}
\caption{Synthetic LTE spectra generated with \textit{Weeds} using the gas parameter in Table \ref{Table:abundances}.}
\label{fig:weeds4}
\end{figure*}

\clearpage
\section{Quantum chemical computations}\label{sec:quant-chem}

We have carried out quantum chemical calculations with the {\sc Gaussian16} \citep[][]{g16} in order to calculate the binding energies of ethylene glycol and formamide on an amorphous water ice surface made of 18 water molecules.
We employed the hybrid-DFT functional BHandHLYP \citep[][]{becke1993,LYP88}, corrected for dispersion with Grimme's G3(BJ) \textit{a posteriori} correction \citep[][]{D3-grimme2010,d3bj_grimme}. Geometry optimisation and frequency calculations were carried out with the 6-311+G(d,p) basis set \citep[][]{hehre_selfconsistent_1972,hariharan_influence_1973, krishnan_selfconsistent_1980}. DFT energies were further refined at the 6-311++G(2df,2pd).
This functional has been shown to provide with high quality binding energies for organic fragments \citep[e.g.][]{Enrique-Romero2019, EnriqueRomero2021, Enrique-Romero2022}.

\begin{figure}[!htbp]
    \centering
    \includegraphics[width=0.8\columnwidth]{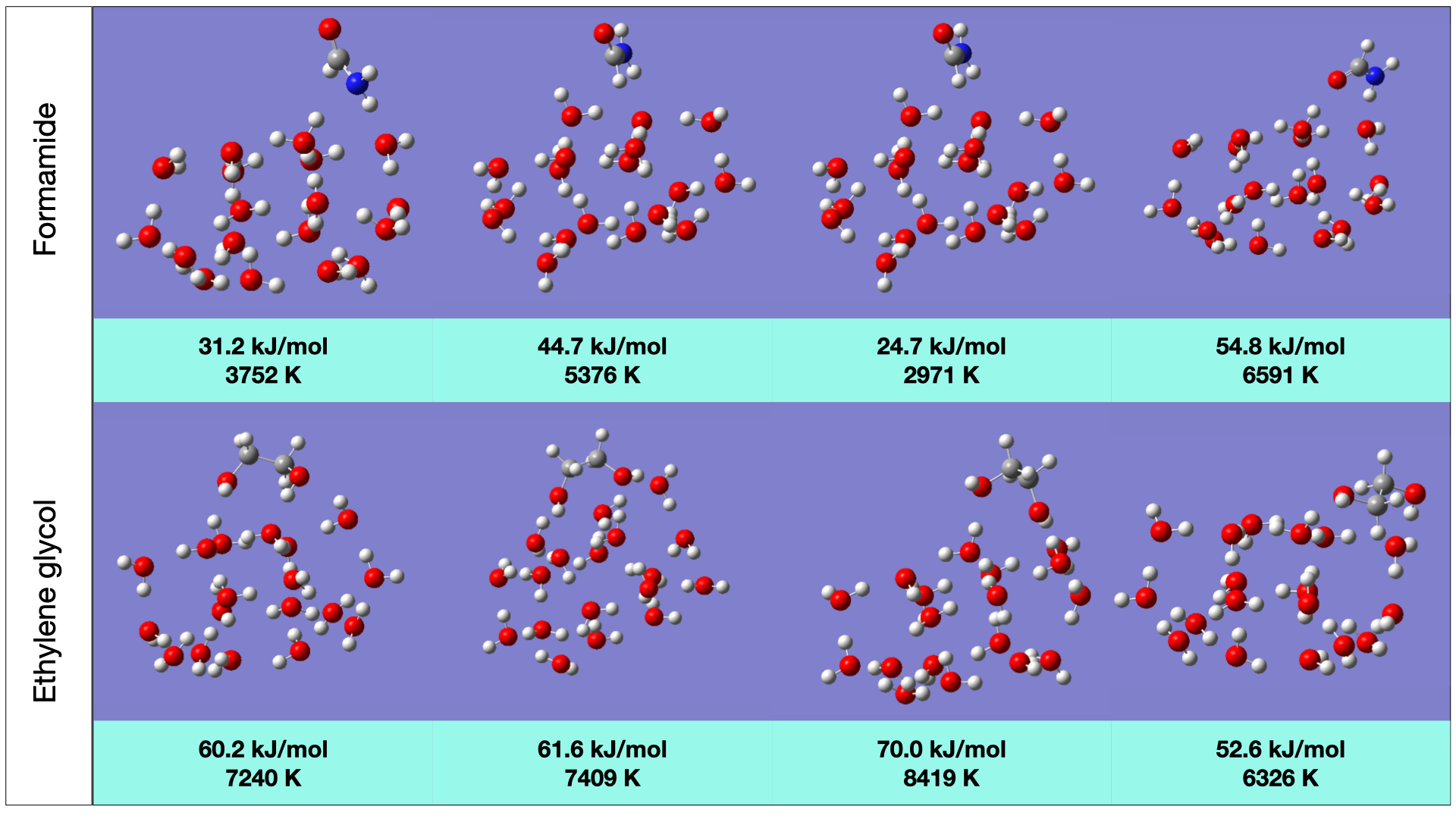}
    \caption{Optimized binding geometries and energies. The energetics include zero point energy corrections.}
    \label{fig:binding_energies}
\end{figure}

Several binding geometries were found for each molecule. For each one of them we provide a binding energy value calculated following $E_{bind} = -(E_{complex}-E_{admol}-E_{surf}) $, where $E_{complex}$, $E_{admol}$, $E_{surf}$ correspond to the zero-point corrected energies of the complex (surface + molecule), the molecule (isolated) and the surface (isolated), respectively. Both, the geometries and their binding energy values are shown Figure \ref{fig:binding_energies}.

As it can be seen, ethylene glycol has a higher binding energy, ranging from $\sim$50--70 kJ/mol (6300--8400 K), than formamide ($\sim$25--55 kJ/mol, 2970--6600K) as a consequence of its flexibility and size, which allow ethylene glycol to establish 2 to 4 H-bonds, while formamide can only make 2--3 at the most, on our water surface model.

\end{document}